\pdfoutput=1

\documentclass[11pt,twoside,a4paper,cmspaper,final,collab]{cms-tdr}
\def\svnVersion{487596P}\def\cmsCernNoTag{CERN-EP-2018-132}\def\cmsCernDate{\today}\def\cmsMessage{Published in Physics Letters B as \href{http://dx.doi.org/10.1016/j.physletb.2019.01.005}{\doi{10.1016/j.physletb.2019.01.005.}}}

\begin{document}\cmsNoteHeader{SUS-17-009}

\hyphenation{had-ron-i-za-tion}
\hyphenation{cal-or-i-me-ter}
\hyphenation{de-vices}
\RCS$Heal: svn+ssh://svn.cern.ch/reps/tdr2/papers/SUS-17-009/trunk/SUS-17-009.tex $
\RCS$Id: SUS-17-009.tex 486561 2019-01-18 13:44:51Z mvesterb $
\newlength\cmsFigWidth
\ifthenelse{\boolean{cms@external}}{\setlength\cmsFigWidth{0.85\columnwidth}}{\setlength\cmsFigWidth{0.4\textwidth}}
\ifthenelse{\boolean{cms@external}}{\providecommand{\cmsLeft}{top\xspace}}{\providecommand{\cmsLeft}{left\xspace}}
\ifthenelse{\boolean{cms@external}}{\providecommand{\cmsRight}{bottom\xspace}}{\providecommand{\cmsRight}{right\xspace}}
\newcommand{\mttwo}{\ensuremath{M_{\text{T2}}}\xspace}
\newlength\cmsTabSkip\setlength{\cmsTabSkip}{1ex}
\newcommand{\lint}{35.9\fbinv}
\newcommand{\mll}{\ensuremath{m_{\ell\ell}}\xspace}
\newcommand{\mlsp}{\ensuremath{m_{\PSGczDo}}\xspace}
\newcommand{\mslep}{\ensuremath{m_{\slep}}\xspace}
\newcommand{\lsp}{\PSGczDo}
\newcommand{\mcha}{\ensuremath{\PSGcpmDo}\xspace}
\newcommand{\mneu}{\ensuremath{\PSGczDt}\xspace}
\newcommand{\MuMu}{\ensuremath{\Pgm^\pm\Pgm^\mp}\xspace}
\newcommand{\ElEl}{\ensuremath{\Pe^\pm\Pe^\mp}\xspace}
\newcommand{\TauTau}{\ensuremath{\tau^\pm\tau^\mp}\xspace}
\newcommand{\EM}{\ensuremath{\Pe^\pm\Pgm^\mp}\xspace}
\newcommand{\rmue}{\ensuremath{r_{\Pgm/\Pe}}\xspace}
\newcommand{\rmuec}{\ensuremath{r_{\Pgm/\Pe,\text{c}}}\xspace}
\newcommand{\Rsfof}{\ensuremath{R_{\text{SF/DF}}}\xspace}
\newcommand{\Reeof}{\ensuremath{R_{\Pe\Pe\text{/DF}}}\xspace}
\newcommand{\Rmmof}{\ensuremath{R_{\Pgm\Pgm\text{/DF}}}\xspace}
\newcommand{\RT}{\ensuremath{R_{\text{T}}}\xspace}
\newcommand{\mt}{\ensuremath{M_{\text{T}}}\xspace}
\newcommand{\mttwol}{\ensuremath{M_{\text{T2}}(\ell\ell)}\xspace}
\newcommand{\mttwolb}{\ensuremath{M_{\text{T2}}(\ell \PQb \ell \PQb)}\xspace}
\newcommand{\slep}{\ensuremath{\widetilde{\ell}}\xspace}
\newcommand{\smuL}{\ensuremath{\widetilde{\mu}_\cmsSymbolFace{L}}\xspace}
\newcommand{\smuR}{\ensuremath{\widetilde{\mu}_\cmsSymbolFace{R}}\xspace}
\newcommand{\stauL}{\ensuremath{\widetilde{\tau}_\cmsSymbolFace{L}}\xspace}
\newcommand{\stauR}{\ensuremath{\widetilde{\tau}_\cmsSymbolFace{R}}\xspace}

\cmsNoteHeader{SUS-17-009}
\title{Search for supersymmetric partners of electrons and muons in proton-proton collisions at $\sqrt{s}=13\TeV$}

\date{\today}

\abstract{
A search for direct production of the supersymmetric (SUSY) partners of electrons or muons is presented in final states with two opposite-charge, same-flavour leptons (electrons and muons), no jets, and large missing transverse momentum. The data sample corresponds to an integrated luminosity of 35.9\fbinv of proton-proton collisions at $\sqrt{s}=13\TeV$, collected with the CMS detector at the LHC in 2016. The search uses the \mttwo variable, which generalises the transverse mass for systems with two invisible objects and provides a discrimination against standard model backgrounds containing $\PW$ bosons. The observed yields are consistent with the expectations from the standard model. The search is interpreted in the context of simplified SUSY models and probes slepton masses up to approximately 290, 400, and 450\GeV, assuming right-handed only, left-handed only, and both right- and left-handed sleptons (mass degenerate selectrons and smuons), and a massless lightest supersymmetric particle. Limits are also set on selectrons and smuons separately. These limits show an improvement on the existing limits of approximately 150\GeV.
}

\hypersetup{
pdfauthor={CMS Collaboration},
pdftitle={Search for supersymmetric partners of electrons and muons in proton-proton collisions at sqrt(s)=13 TeV},
pdfsubject={CMS},
pdfkeywords={CMS, physics, SUSY}}

\maketitle
\section{Introduction}
\label{sec:introduction}

The standard model (SM) of particle physics provides a description of the fundamental particles and their interactions, and its predictions have been confirmed experimentally with increasing
precision over the last several decades. Supersymmetry (SUSY)~\cite{Ramond:1971gb,Golfand:1971iw,Neveu:1971rx,Volkov:1972jx,Wess:1973kz,Wess:1974tw,Fayet:1974pd,Nilles:1983ge}, one of the most
promising extensions of the SM, addresses several open questions for which the SM has no answer, such as the hierarchy problem and the origin of dark matter. The theory postulates a new
fundamental symmetry that assigns to each SM particle a SUSY partner whose spin differs by one half, causing the SUSY partner of an SM fermion (boson) to be a boson (fermion). In addition to
stabilising the Higgs boson ($\PH$) mass via cancellations between quantum loop corrections including the top quark and its superpartner, SUSY provides
a natural dark matter candidate, if $R$-parity~\cite{Farrar:1978xj} is conserved, in the form of the lightest SUSY particle (LSP), which is assumed to be massive and stable.

SUSY particles (sparticles) that are coloured, the squarks and gluinos, are produced via the strong interaction with significantly larger cross sections than colourless sparticles of equal masses, at the Large Hadron Collider (LHC).
However, if the squarks and gluinos are too heavy to be produced at the LHC, the direct production of colourless sparticles, such as the electroweak superpartners
(charginos ($\mcha$), neutralinos ($\mneu$), and sleptons ($\slep$)), would be the dominant observable SUSY process.

\begin{figure}[!htb]
\centering
\includegraphics[width=0.5\textwidth]{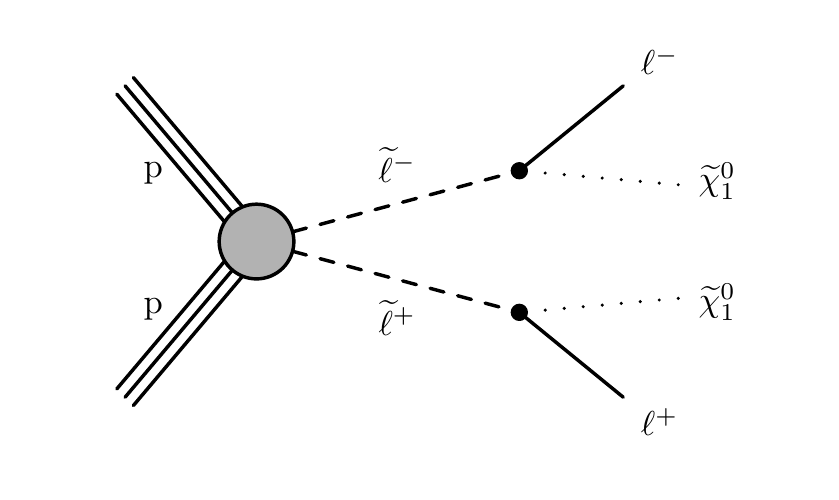}
\caption{
\label{fig:diagram}
Diagram of slepton pair production with direct decays into leptons and the lightest neutralino. }
\end{figure}

Supersymmetric models predict charged sleptons (
$\seL$, $\smuL$, $\stauL$, $\seR$, $\smuR$, $\stauR$), the superpartners of the charged left-handed and right-handed SM leptons, which can be produced at proton-proton ($\Pp\Pp$) colliders in direct
electroweak pair production. At sufficiently heavy slepton masses, the sleptons undergo a two-body decay into one of the heavier neutralinos
or a chargino, while direct decays to a neutralino LSP are favoured for light slepton masses.
This Letter presents a search for directly produced selectrons and smuons ($\seL$, $\smuL$, $\seR$, $\smuR$),
under the assumption of direct decays $\slep\to \ell\lsp$ with 100\% branching ratio, as sketched
in Fig.~\ref{fig:diagram}. The final state contains little or no hadronic activity and provides a clean signature
composed of two opposite-charge (OC), same-flavour (SF) leptons (dielectron or dimuon pairs) and large
missing transverse momentum (\ptmiss) from the two LSPs that escape detection.

The main SM backgrounds resulting in two OC SF leptons and no reconstructed jets are $\Pp\Pp\to\ttbar$ (if both jets from the top decays are out of acceptance) and
$\Pp\Pp\to\PW\PW \to 2\ell 2\nu$, both of which involve \PW\ bosons that decay into an electron or
a muon with equal probability, resulting in the same number of dielectron and dimuon events as electron-muon events (different flavour, DF).
This flavour symmetry is used in the analysis to predict the number of background SF leptons based on the number of DF leptons in the signal region (SR) in data, after correcting for differences in
trigger and lepton reconstruction efficiencies.
The Drell--Yan (DY) process would also be a main background in the analysis, but is greatly suppressed by the SR requirements.
The $\Pp\Pp\to\PZ\PZ \to 2\ell 2\nu$ and $\Pp\Pp\to\PW\PZ\to3\ell\nu$ processes can also
result in two OC SF leptons.
These contributions are taken from Monte Carlo (MC) simulation after comparing data and simulation predictions in control regions (CR).

The data set of proton-proton collisions used for this search was collected in 2016 with the CMS detector
at a centre-of-mass energy of $\sqrt{s}=13\TeV$, and corresponds to an integrated luminosity of \lint.
Interpretations of the search results are given in terms of simplified SUSY model spectra~\cite{Alves:2011wf,Chatrchyan:2013sza}.
Searches for SUSY in these final states were performed previously by the
ATLAS~\cite{Aad:2014vma} and CMS~\cite{2012ewk} Collaborations at $\sqrt{s}=8\TeV$, by the ATLAS~\cite{Aaboud:2018jiw} Collaboration at $\sqrt{s}=13\TeV$,
and a complementary search targeting scenarios where the mass difference between the LSP and the slepton is small
has been performed by the ATLAS Collaboration~\cite{Aaboud:2017leg} at $\sqrt{s}=13\TeV$.
\section{The CMS detector}
\label{sec:cmsdetector}

The central feature of the CMS apparatus is a superconducting solenoid, 13\unit{m} in length and 6\unit{m} in diameter, that provides
an axial magnetic field of 3.8\unit{T}. Within the solenoid volume are various particle detection systems. Charged-particle
trajectories are measured by silicon pixel and strip trackers, covering $0<\phi\leq2\pi$ in azimuth and $\abs{\eta}<2.5$,
where the pseudorapidity $\eta$ is defined as $-\log [\tan(\theta/2)]$, with $\theta$ being the polar angle of the
trajectory of the particle with respect to the counterclockwise-beam direction.
A lead tungstate crystal electromagnetic calorimeter (ECAL), and a brass and scintillator hadron calorimeter surround the tracking volume.
The calorimeters provide energy and direction measurements of electrons and hadronic jets.
Muons are detected in gas-ionisation detectors embedded in the steel flux-return yoke outside the solenoid.
The detector is nearly hermetic, allowing for transverse momentum (\pt) balance measurements, in the plane perpendicular to the beam direction.
A two-tier trigger system selects events of interest for physics analysis.
A more detailed description of the CMS detector, together with a definition of the coordinate system used
and the relevant kinematic variables, can be found in Ref.~\cite{Chatrchyan:2008zzk}.

\section{Event samples}
\label{sec:samplesObjects}
The search is based on samples of dielectron and dimuon events.  As mentioned in Section~\ref{sec:introduction},
DF events are used to predict the contribution of background SF events in the SR.
The SF and DF samples are collected with a variety of isolated and non-isolated dilepton triggers.
Triggers that include loose isolation criteria on both leptons require $\pt>23\GeV$ (electron) or 17\GeV (muon) on the highest \pt lepton.
The other lepton is then required to have $\pt>12\GeV$ (electrons) or 8\GeV (muons).
In addition, dilepton triggers without isolation requirements are used to increase the signal efficiency.
These require $\pt>33$ (30)\GeV for both leptons in the dielectron (electron-muon) case.
The dimuon trigger requires either $\pt>27$ (8)\GeV for the highest (next-to-highest) \pt muon during early data taking periods, with an increase of the thresholds to $\pt>30$ (11)\GeV for the
highest (next-to-highest) \pt muon during remaining data taking periods. The data collected with these triggers are used for the data-driven background prediction as well as to collect the
events in the SR with a higher leading lepton \pt requirement of 50\GeV.
The lepton pseudorapidity coverage for the triggers is $\abs{\eta}<2.5$ (2.4) for electrons (muons).
The trigger efficiencies are measured in data using events selected by a suite of jet triggers and are found to be 90--96\%.

The main SM backgrounds are estimated using data control samples, while simulated events are used
to predict backgrounds from diboson (\cPZ\cPZ\ and \PW\cPZ) production. Simulated events are
also used extensively in the analysis to estimate systematic uncertainties.
Next-to-leading order (NLO) and next-to-NLO (NNLO) cross
sections~\cite{Alwall:2014hca,Czakon:2011xx,Grazzini:2017ckn,Cascioli:2014yka,Caola:2015psa,Campbell:2016ivq,Binoth:2008kt,Nhung:2013jta,Yong-Bai:2015xna,Hong:2016aek,Yong-Bai:2016sal,Dittmaier:2017bnh}
are used to normalise the simulated background samples.
For the signal samples we use NLO plus next-to-leading-logarithmic (NLL) calculations for left- or right-handed sleptons,
with all the other sparticles except the LSP assumed to be heavy and decoupled~\cite{Beenakker:1999xh,Fuks:2012qx,Fuks:2013vua}.

The $\Pg\Pg\to\PZ\PZ$ process is generated at LO with \MCFM~7.0~\cite{Campbell:2010ff}, and all other diboson production processes~\cite{Melia:2011tj,Nason:2013ydw}, and \ttbar~\cite{Campbell:2014kua}
and the production of single top quark associated with a \PW\ boson~\cite{Re:2010bp}, are generated at NLO with no additional partons with {\POWHEG}~v2.
Simulated samples of DY processes are generated with \MGvATNLO 2.3.3 program~\cite{Alwall:2014hca} to leading order precision with up to four additional partons in the matrix element
calculation. Simulated VVV and $\ttbar$V (V$\,=\PW,\PZ$) events are simulated with the same generator but at NLO precision.
The NNPDF3.0~\cite{Ball:2014uwa} LO (NLO) parton distribution functions (PDFs) are used for
the samples generated at LO (NLO).
The matrix element calculations performed with these generators are interfaced with \PYTHIA~\cite{Sjostrand:2007gs},
including the CUETP8M1 tune~\cite{Skands:2014pea,CMS-PAS-GEN-14-001} for the simulation of parton showering and hadronisation.
Double counting of partons generated
with \MGvATNLO and \PYTHIA is removed using the MLM~\cite{Alwall:2007fs} and {\sc FxFx}~\cite{Frederix:2012ps}
matching schemes in the LO and NLO samples, respectively.
The detector response is simulated with a \GEANTfour model~\cite{Geant} of the CMS detector.
The simulation of new-physics signals is performed using the \MGvATNLO 2.2.2 program at LO precision, with up to
two additional partons in the matrix element calculation. Events are then interfaced with \PYTHIA for fragmentation
and hadronisation and simulated using the CMS fast simulation package~\cite{fastsim}. The slepton decays are also simulated with \PYTHIA.
Multiple \Pp\Pp\ interactions, also known as pileup, are superimposed on the hard
collision, and the simulated samples are reweighed in such a way that the number of collisions per bunch crossing
accurately reflects the distribution observed in data.
Corrections are applied to the simulated samples to
account for differences between simulation and data in the trigger and reconstruction efficiencies.

\section{Object selection}
\label{sec:samplesObjects2}
The particle-flow (PF) algorithm~\cite{Sirunyan:2017ulk} reconstructs and identifies particle candidates in the event,
referred to as PF objects. To select collision events we require at least one reconstructed vertex, and the one with the largest value of summed physics object $\pt^{2}$ is taken to be the
primary $\Pp\Pp$ interaction vertex.
The physics objects used for the primary vertex selection are the objects returned by a jet finding algorithm~\cite{Cacciari:2008gp,FastJet} applied to all charged tracks
associated with the vertex, plus the corresponding associated \ptmiss. Its vector \ptvecmiss is defined as the projection onto the plane perpendicular to the beam axis
of the negative vector sum of the momenta of all reconstructed PF objects in the event, and its magnitude is \ptmiss.
Electrons are reconstructed by associating tracks with ECAL clusters. They are identified using a multivariate approach based on information on
ECAL cluster shapes, track reconstruction quality, and the matching between the track and the ECAL cluster~\cite{Khachatryan:2015hwa}.
Electrons coming from reconstructed photon conversions are rejected.
Muons are reconstructed from tracks in the muon system associated with tracks in the tracker.
The identification uses the quality of the track fit and the number of associated hits in the tracking detectors~\cite{Sirunyan:2018fpa}.
For both electrons and muons, the impact parameter with respect to the primary vertex is required to be within 0.5\unit{mm} in the transverse
plane and less than 1\unit{mm} along the beam direction.
A lepton isolation variable is defined as the scalar \pt\ sum
of all PF objects in a cone around the lepton, excluding identified electrons or muons.
The effect of additional \Pp\Pp\ interactions in the same or nearby bunch crossings (pileup) can be mitigated
by only considering charged PF objects that are compatible with the primary vertex and the per-event average expected pileup contribution is subtracted from the neutral component of the isolation.
The isolation sum is required to be smaller than 10 (20)\% of the electron (muon) \pt.
A shrinking cone-size with increasing \pt is chosen that ensures high efficiency for leptons from Lorentz-boosted boson decays~\cite{Rehermann:2010vq}.
This varying cone size is chosen as the following $\DR=\sqrt{\smash[b]{(\Delta\phi)^2+(\Delta\eta)^2}}=0.2$ for $\pt<50\GeV$, $=10\GeV/\pt$ for $50<\pt<200\GeV$,
and $0.05$ for $\pt > 200\GeV$.

Isolated charged particle tracks identified by the PF algorithm are selected with a looser criteria than the leptons defined above,
and are used as a veto on the presence of additional charged leptons from vector boson decays.
Isolation is evaluated by summing the \pt of all charged PF objects within a cone
of $\DR=0.3$ and with the longitudinal impact parameter $\abs{\Delta z}<1$\unit{mm} relative to the primary vertex.
PF objects identified as charged hadrons (electrons or muons) are required to have $\pt>10$ (5)\GeV and an isolation value less than 10 (20)\% of the object \pt.

Jets are clustered from PF objects, excluding charged hadrons not associated with
the primary vertex that are assumed to be the result of pileup interactions,
using the anti-\kt\ clustering algorithm~\cite{Cacciari:2008gp} with a distance parameter of 0.4 as
implemented in the \FASTJET\ package~\cite{FastJet,Cacciari:2005hq}.
Jets are required to have $\abs{\eta}<2.4$ and $\pt>25\GeV$,
where the \pt is corrected for non-uniform detector response
and pileup effects~\cite{1748-0221-6-11-P11002,cacciari-2008-659}.
Jets reconstructed within $\DR < 0.4$ of any of the selected leptons are removed from the event.
Corrections to the jet energy are propagated to \ptmiss\ using the procedure developed in Ref.~\cite{1748-0221-6-11-P11002}.
At least two jets of \pt above 35\GeV are selected for CRs of this analysis, and events are vetoed that contain jets with \pt above 25\GeV in the SR.

Events are selected for the SR by requiring two OC SF leptons (\ElEl or \MuMu)
with $\pt>50$ (20)\GeV for the highest (next-to-highest) \pt\ lepton
and $\abs{\eta}<2.4$ for both leptons.
For the background prediction methods a sample of lepton pairs is selected, with a \pt threshold of 25 (20)\GeV for the leading (subleading) lepton.
The highest minimum \pt value is chosen because it efficiently suppresses backgrounds while maintaining signal acceptance efficiency.
Events with additional leptons, identified with the looser requirement that the isolation sum should be less than 40\% of the lepton \pt, are vetoed.
Leptons must be spatially separated by $\DR > 0.1$ to avoid reconstruction efficiency differences between electrons and muons in events with collinear leptons.
All events containing leptons in the transition region between the barrel and endcap of the ECAL,
$1.4<\abs{\eta}<1.6$, are rejected to ensure similar acceptance for electrons and muons.
The same lepton selection criteria are used for a control sample of OC DF pairs, \EM.
The selection requirements have been chosen in order to maximise the lepton selection efficiency
while maintaining a similarity between electron and muon efficiencies.

\section{Search strategy}
\label{subsec:signalregions}

The slepton SRs are designed to suppress expected backgrounds from SM processes, while maintaining sensitivity
to different assumptions on the masses of the \slep and \lsp.
To suppress backgrounds due to low-mass resonances and \cPZ\ boson production, the dilepton invariant mass is required
to be above 20\GeV, and to be either below 76 or above 106\GeV.
Little or no hadronic activity is expected
in the direct production of sleptons at \Pp\Pp\ colliders when assuming a 100\% branching ratio for $\slep\to\ell\lsp$.
As a result, events are rejected if they contain jets with \pt above 25\GeV.
Furthermore, events with two leptons and an additional isolated and charged PF candidate passing the selections described
in Section~\ref{sec:samplesObjects2} are vetoed in order to reduce the background from events with more than two isolated leptons.

The kinematic variable \mttwo~\cite{MT2variable,MT2variable2} is used to reduce backgrounds from \ttbar and \PW\PW\ processes.
This variable was first introduced to measure the mass of pair-produced particles, each decaying to the same final state,
consisting of a visible and an invisible particle.
It is defined as:
\begin{linenomath}
\begin{equation}
\begin{aligned}
\mttwo = \min_{\ptvecmiss{}^{(1)} + \ptvecmiss{}^{(2)} = \ptvecmiss} \left[ \max \left( \mt^{(1)} , \mt^{(2)} \right) \right],
\end{aligned}
\end{equation}
\label{eq:MT2}
\end{linenomath}
where $\ptvecmiss{}^{(i)}$ ($i=1,2$) are trial vectors obtained by decomposing \ptvecmiss.
The transverse masses $\mt^{(i)} = \sqrt{\smash[b]{2 \pt^{\text{vis}} \ptmiss{}^{(i)} [1 - \cos(\Delta\phi)]}}$ are obtained by pairing either of these trial vectors with one of the two leptons.

The $\Delta\phi$ is the angle between the \pt of the lepton (noted as $\pt^{\text{vis}}$) and $\ptvecmiss{}^{(i)}$.
The minimisation is performed over all trial momenta satisfying the \ptvecmiss\ constraint.
When building \mttwo from the two selected leptons and \ptvecmiss, denoted as \mttwol, its distribution exhibits a sharp
decrease above the mass of the \PW\ boson for \ttbar\ and \PW\PW\ events and is therefore well suited to suppress these backgrounds.
For this reason a requirement of $\mttwol>90\GeV$ is imposed in this search.

The SR is divided into four bins of \ptmiss: 100--150, 150--225, 225--300, and $\geq$300\GeV. The selection results in a signal selection efficiency that ranges from  20 to 30\% assuming a massless
LSP.
The simplified models do not assume that smuon and selectron masses should be the same, so the results are presented for dielectron and dimuon pairs separately.
Since the search for combined SF dilepton pairs (i.e. dielectrons $+$ dimuons) is able to employ additional background estimation techniques which lower the overall background uncertainty, the corresponding results are also quoted separately.

\section{Standard model background predictions}
\label{sec:backgrounds}

The backgrounds from the SM processes are divided into four categories.
Flavour-symmetric (FS) background processes are those processes that result in DF pairs (\EM) as often as SF pairs (\MuMu, \ElEl).
The dominant contributions to this category are due to top quark pair production and $\PW\PW$ production, but also processes such as $\PZ\to\TauTau$ are estimated with this method.

Diboson production, $\PZ\PZ$ and $\PW\PZ$, can yield OC SF leptons, and this contribution is estimated from simulation.
The $\PZ\PZ$ process can result in  a final state with two leptons originating from one $\PZ$ boson decay and two neutrinos
from the other $\PZ$ boson decay.
The $\PW\PZ$ process can give rise to a final state with three leptons and \ptmiss,
which can satisfy the signal selection criteria if one of the leptons fails the identification
or acceptance requirements.

The contribution from DY ($\PZ\to\ElEl$ and $\PZ\to\MuMu$) is small in the SR due to the large \ptmiss requirement. The contribution is estimated using simulated events after
relaxing the $\PZ$ boson veto and a transfer factor $r_{\text{out/in}}$ that gives the contribution of DY events outside of the $\PZ$ boson mass window of 76--106\GeV.
Furthermore, leptons from $\PZ\to\TauTau$ decays are vetoed, as this background is FS.
The transfer factor $r_{\text{out/in}}$ is measured in a DY enriched CR as the ratio of events outside of the $\PZ$ boson mass over the events compatible with the $\PZ$ boson mass.
From simulation studies a systematic uncertainty of 50\% is found to cover any dependencies of the transfer factor $r_{\text{out/in}}$ on the \ptmiss and the \mttwo, and is
assigned to the method.

Finally, a very minor background, referred to in the following as Rare backgrounds, originates from triboson production, or processes resulting in non-FS leptons,
such as $\ttbar\PZ$, $\cPqt\PZ\Pq$ and $\cPqt\PW\PZ$.
The simulation is also used to estimate this contribution, with a conservative systematic uncertainty of 50\% assigned in place of QCD scale and PDF variations.
\subsection{Flavour-symmetric backgrounds}
\label{sub:fsbkg}

This paper presents limits on the direct production of sleptons, selectrons and smuons in the SF, dielectron and dimuon final states. For the results in the dielectron and dimuon final states,
the SM dielectron ($N_{\EE}$) and dimuon ($N_{\MM}$) backgrounds are obtained using event counts in the DF sample ($N_{DF}$) multiplied by a translation factor \Reeof and \Rmmof respectively,
according to
\begin{linenomath}
\begin{equation}
N_{\EE}  = \Reeof \times N_{DF},  N_{\MM}  = \Rmmof \times N_{DF}.
\end{equation}
\label{eq:rEEMMOF}
\end{linenomath}
For the results in the SF final state, prediction of SF backgrounds ($N_{SF}$) is similarly obtained using event counts in the DF sample ($N_{DF}$), multiplied by a translation factor, \Rsfof,
according to
\begin{linenomath}
\begin{equation}
N_{SF}  = \Rsfof \times N_{DF}.
\end{equation}
\label{eq:rSFOF}
\end{linenomath}
The translation factors \Reeof and \Rmmof are estimated through a measurement of the rate of dielectron and dimuon events to DF events in a dedicated CR.
The translation factor \Rsfof is measured, similarly to the \Reeof and \Rmmof, as the rate of SF events to DF events in a dedicated CR.
Another method to estimate the SF yields is measuring the difference for electrons and muons in reconstruction, identification and trigger efficiencies.
As the second method uses information of both electrons and muons, it cannot be used for the estimation of the \Reeof and \Rmmof.
However, it is combined with the results from the initial measurement of \Rsfof using the weighted average according to their uncertainties as described in Ref.~\cite{Sirunyan:2017qaj}, and results in a reduction in the systematic uncertainty that comes from the combination of the two methods.
The first method estimates directly the translation factors \Reeof, \Rmmof and \Rsfof in a data CR enriched in \ttbar events, requiring exactly two jets, $100<\ptmiss<150\GeV$, and excluding the
dilepton invariant mass range $70<\mll<110\GeV$ to reduce contributions from DY production.
The \Rsfof, \Reeof and \Rmmof are computed using the observed yield of the SF, dielectron and dimuon events compared to the observed yield of DF events, $\Rsfof = N_{\text{SF}}/N_{\text{DF}}$,
$\Reeof = N_{\text{\EE}}/N_{\text{DF}}$ and $\Rmmof = N_{\text{\MM}}/N_{\text{DF}}$ respectively. Data and simulation agree within 2\% in this region.
A 4\% systematic uncertainty on the translation factor is assigned from simulation studies, as the maximal magnitude of the systematic needed to cover discrepancies in the translation factor
as a function of some SR variables. The main SF backgrounds estimated with the method described above are \ttbar and $\PW\PW$. Simulation studies show that the $\PW\PW$ is the dominating FS
process at high \ptmiss and that there is no dependence on the \Rsfof, \Reeof and \Rmmof factors arising from the different processes.

The second method utilises a factorised approach. The ratio of muon to electron reconstruction and identification efficiencies, \rmue, is measured in a CR enriched in DY events by requiring at
least two jets, $\ptmiss<50\GeV$, and $60<\mll<120\GeV$.
Assuming factorisation for the efficiencies of the two leptons, the ratio of efficiencies for muons and electrons is measured as $\rmue = \sqrt{\smash[b]{N_{\MM}/N_{\EE}}}$.
This ratio depends on the lepton \pt due to the trigger and reconstruction efficiency differences, especially at low lepton \pt,
and a parametrisation as a function of the \pt of the less energetic lepton is used:
\begin{linenomath}
\begin{equation}
\rmue  = \rmuec +  \frac{\alpha}{\pt}.
\end{equation}
\label{eq:rMuEFormular}
\end{linenomath}
Here \rmuec and $\alpha$ are constants that are determined from a fit to data and cross-checked using simulation.
These fit parameters are determined to be $\rmuec=1.140\pm0.005$ and $\alpha=5.20\pm0.16\GeV$.
In addition to the fit uncertainty, a 10\% systematic uncertainty is assigned to account for variations observed when studying the dependence of \rmue on \ptmiss and on the \pt of the more energetic lepton.

The trigger efficiencies for the three flavour combinations are used to define the factor
$\RT = \sqrt{\smash[b]{\epsilon^{\text{T}}_{\MuMu}\epsilon^{\text{T}}_{\ElEl}}}/\epsilon^{\text{T}}_{\Pe^{\pm}\Pgm^{\mp}}$,
which takes into account the difference between SF and DF channels.

The efficiencies, $\epsilon^{\text{T}}_{\MuMu}$, $\epsilon^{\text{T}}_{\ElEl}$ and $\epsilon^{\text{T}}_{\Pe^{\pm}\Pgm^{\mp}}$,
are calculated as the fraction of events in a control sample recorded with non-leptonic triggers that would also pass the dimuon, dielectron and electron-muon trigger selection, respectively.
The efficiencies are measured to range between 90--96\% depending on the flavour composition of the dilepton trigger, and a systematic uncertainty of 3\% is assigned to each trigger efficiency, which is the maximal deviation between the efficiencies in data and MC.
This results in the final value of $\RT=1.052\pm0.043$, where the uncertainty is due to the error propagation of the uncertainties on the individual efficiencies to \RT.

The factorised approach measures \Rsfof according to $\Rsfof = 0.5(r_{\Pgm\mathrm{/\Pe}}+r_{\Pgm\mathrm{/\Pe}}^{-1}) \RT$ where the factor of 0.5 is due to the assumption
that the number of produced DF events is twice the number of produced events in each SF sample (\Pe\Pe~and $\mu\mu$).
The summation of the \rmue with its inverse leads to a reduction in the associated uncertainty.
As the parameterisation of \rmue in the factorised approach has to be applied on an event-by-event basis, no constant result for \Rsfof can be given.
However, the \Rsfof from the first method can be compared to the results from the second method by estimating the \Rsfof through dividing the number of predicted SF events by the observed DF events in each SR.
Both factors range from 1.08 to 1.1 over all SRs and since the predictions from the two methods agree well they are combined using a weighted average.

\subsection{Diboson backgrounds}
\label{sub:dibosonBkgs}
\begin{figure*}[ht!]
\centering
\includegraphics[width=0.48\textwidth]{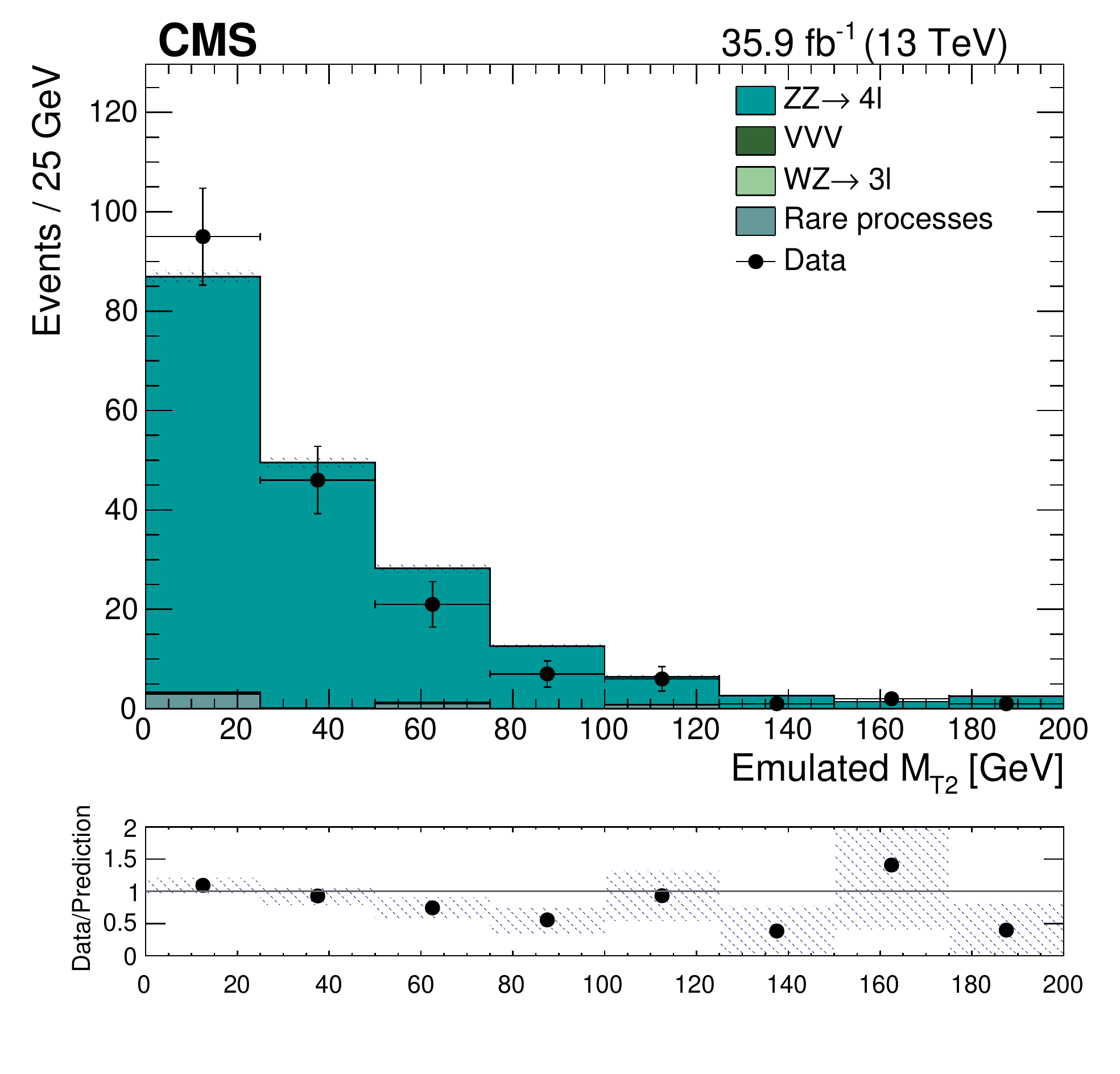}
\includegraphics[width=0.48\textwidth]{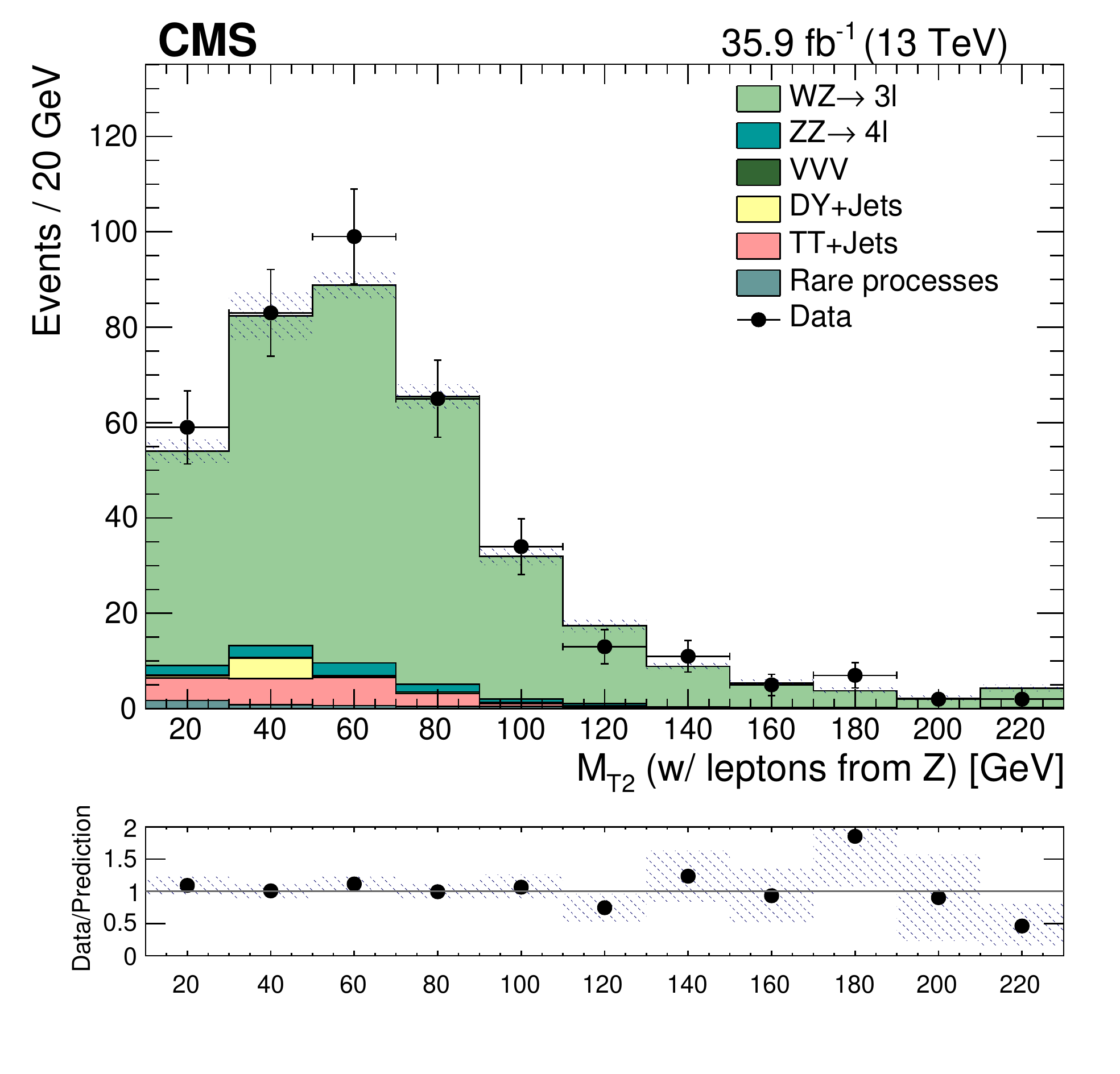}
\caption{Upper: Distribution of \mttwo for the $\PZ\PZ$ (left) and $\PW\PZ$ (right) CRs, in simulation (coloured histograms) and with the corresponding event counts observed in data
(black points). Lower: Ratio of data to simulation, with the filled band representing the statistical uncertainty on the data and the simulations.}
\label{fig:mt2CRs}
\end{figure*}

Although a $\PZ$ boson veto is applied, the $\PZ\PZ$ process can still enter the SR through an off-shell $\PZ$ boson.
This contribution is estimated from simulated events, validated in a data CR with four identified leptons.
The selections for the CR and SR are exclusive, and the
physics process in the CR ($\PZ\PZ$ where both $\PZ$ bosons decay to charged leptons) has similar kinematics
as the process it is designed to validate.
In order for the CR to accurately reflect the kinematics in the SR, the same jet veto as in the SR is applied in the CR.
In addition, for the CR the $\PZ$ boson candidate with the invariant mass best (next best) compatible with the $\PZ$ boson mass is required to have $76 < \mll < 106\GeV$ ($50 < \mll < 130\GeV$).
A generator-level \pt dependent NNLO/NLO K factor of 1.1--1.3, taking into account missing electroweak
corrections~\cite{Bierweiler:2013dja,Gieseke:2014gka,Baglio:2013toa}, is applied to the $\qqbar\to\cPZ\cPZ$ process cross-sections.
The smaller contribution from the $\Pg\Pg\to\PZ\PZ$ process is normalised to the NLO calculation~\cite{Caola:2015psa}.
After subtracting contributions to the CR from other processes, as determined by simulation, a simulation-to-data scale factor of $0.94\pm0.07$ is obtained.
This scale factor is used to correct the $\PZ\PZ$ background prediction from simulation in the SR, where one $\PZ$ boson decays to charged leptons, and the other $\PZ$ boson decays to neutrinos.
A systematic uncertainty of 7\% results from the limited number of events in the CR.
The distribution of \mttwo in the $\PZ\PZ$ CR is shown in Fig.~\ref{fig:mt2CRs}, where the \pt of the two leptons most compatible with the $\PZ$ boson is added to the \ptmiss and the other two leptons are used to form the \mttwo, and show a good agreement between data and simulation.

A difference in the \ptmiss and \mttwol distributions is observed after applying the $\qqbar\to\PZ\PZ$ NNLO/NLO K factor as a function of different generator level kinematic variables.
An uncertainty is then assigned to the method based on the difference in the \ptmiss shape for MC events in the SR before and after the application of the K factor.
Additional systematic uncertainties are considered in the background prediction, originating from the jet energy scale, the variation of the renormalisation and factorisation scales,
the PDF choice, and the uncertainties in the lepton reconstruction and isolation efficiencies, and in the trigger modelling.

The $\PW\PZ$ process result in SF events when one of the three leptons is not reconstructed (lost).
The two detected leptons are of the SF when the lepton from \PW\ decay is lost, but they can be either SF or DF, with equal probability, when the lost lepton is from the \cPZ\ boson decay.
In the first case the background contribution is estimated from simulation, whereas in the second case it is covered by the data-driven FS prediction method.

Just as for the $\PZ\PZ$ background, the prediction from simulation is validated in a CR enriched in $\PW\PZ$ events.
We select events with three leptons, the same jet veto as applied in the SR and a requirement of $\ptmiss>70\GeV$.
The invariant mass of the two SF leptons must be within $76<\mll<106\GeV$.
To increase the purity of the $\PW\PZ$, events are required to have $\mt>50\GeV$, where \mt is calculated from \ptmiss and the lepton from the \PW\ boson.
The distribution of \mttwo in the $\PW\PZ$ CR is shown in Fig.~\ref{fig:mt2CRs}, where the \mttwo is constructed with the two leptons compatible with the $\PZ$ boson and show a good agreement between data and simulation.
After subtracting contributions from other processes, a simulation-to-data scale factor of 1.06 with a systematic uncertainty of 6\% resulting from the limited number of events in the CR is obtained and applied to the prediction from simulation in the SR.
An additional uncertainty of 5\% is added (in quadrature) to cover possible differences in the
identification and isolation efficiencies between data and simulation in the third lepton low \pt region.
Finally, uncertainties due to the jet energy scale, the lepton efficiencies, the trigger modelling, the PDF choice, and the renormalisation and factorisation scales
are taken into account when computing the expected $\PW\PZ$ yields in the SR.

\section{Results}
\label{sec:results}
\begin{figure*}[ht!]
\centering
\includegraphics[width=0.48\textwidth]{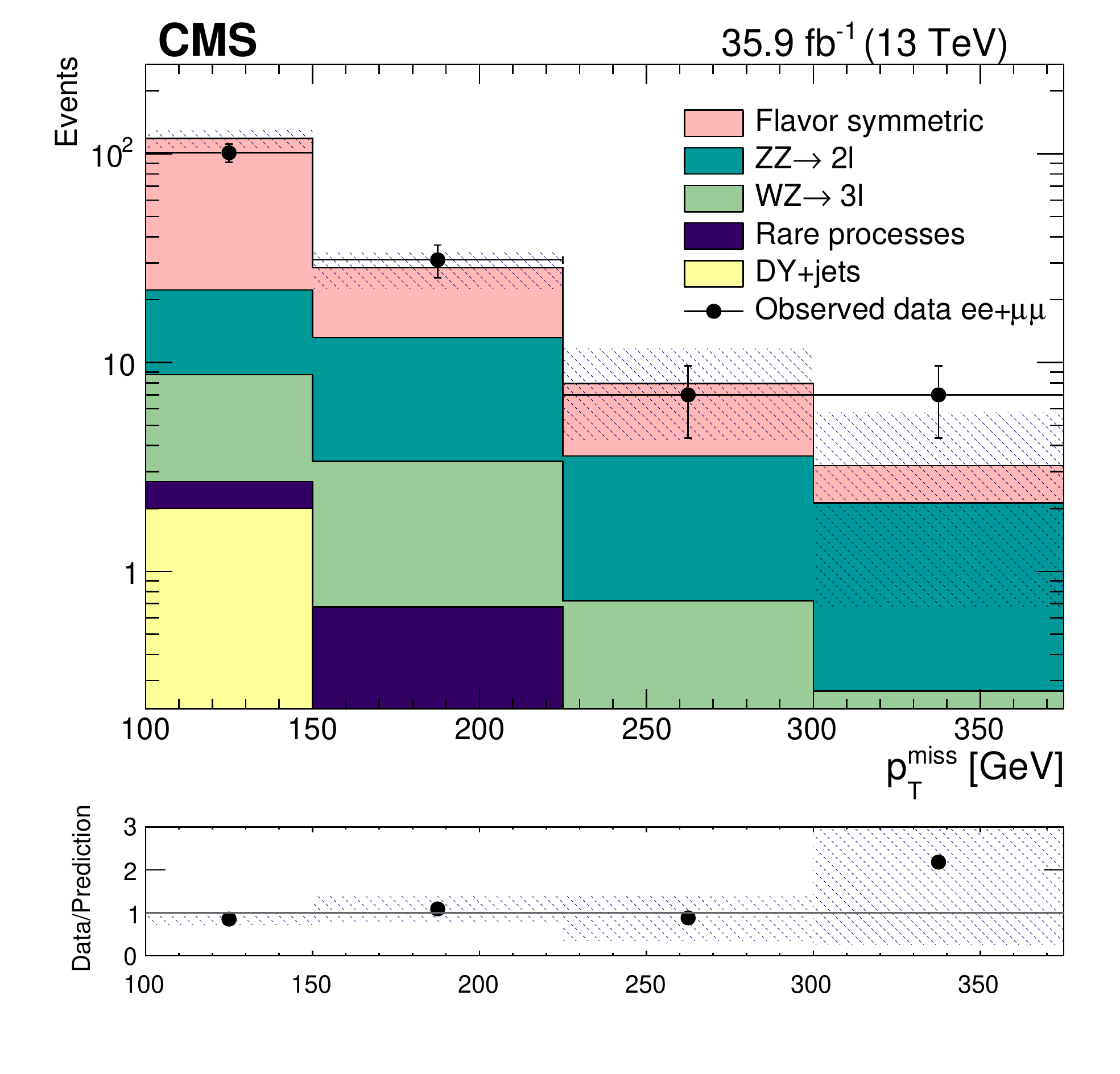}
\caption{Upper: Distribution of \ptmiss for the resulting SM background yields estimated in the analysis SR (coloured histograms) with the corresponding event counts observed in data
(black points), selecting only SF events. Lower: Ratio of data to SM prediction, with the filled band representing the statistical uncertainty on the data and
the estimated backgrounds and the systematic uncertainty on the estimated backgrounds.}
\label{fig:results}
\end{figure*}

\begin{figure*}[ht!]
\centering
\includegraphics[width=0.48\textwidth]{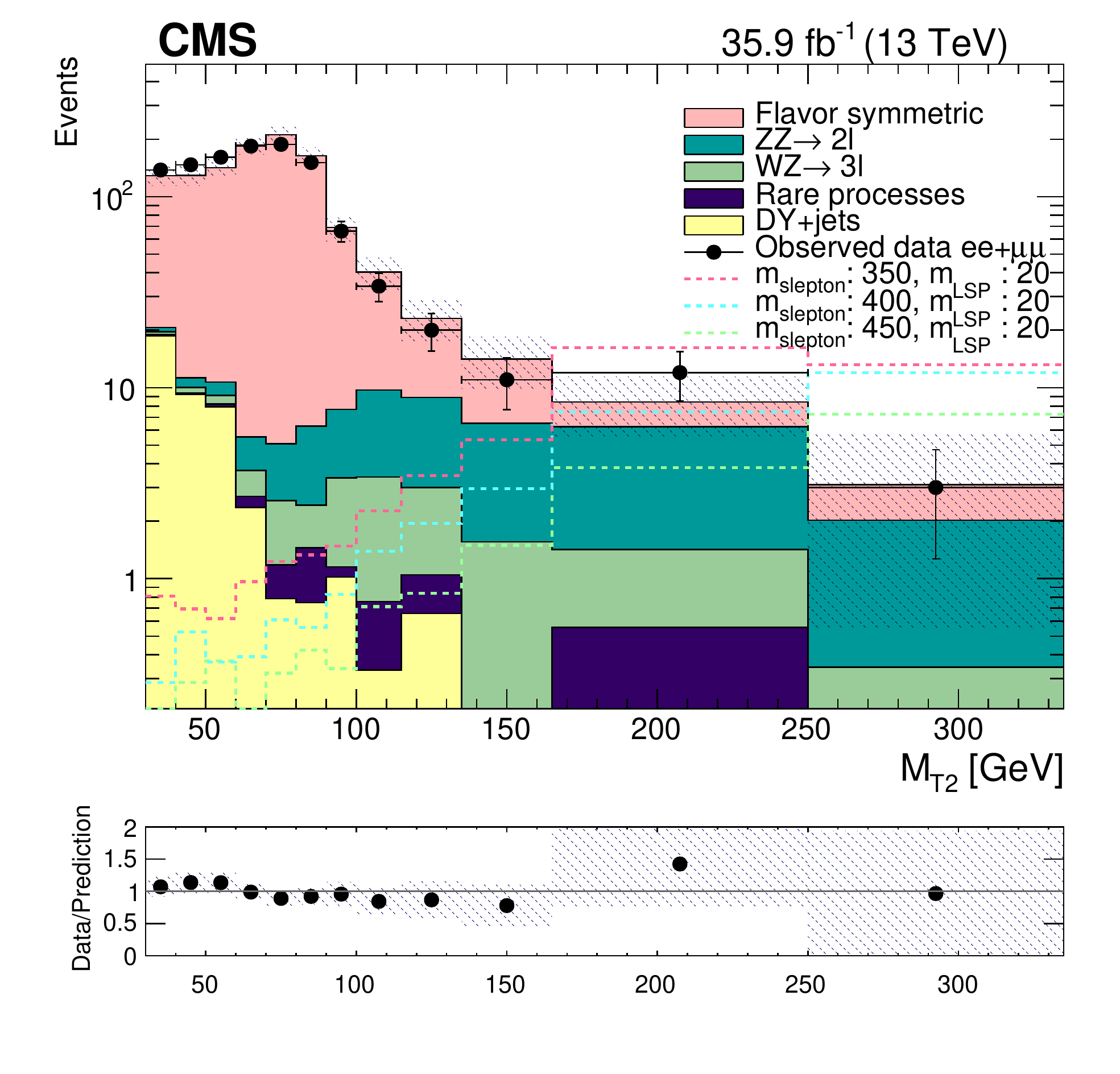}
\caption{Upper: Distribution of \mttwo for the resulting SM background yields estimated in the analysis SR (coloured histograms) with the corresponding event counts observed in data
(black points), and three signal scenarios (hatched lines), selecting only SF events and assuming the production of both left and right-handed sleptons.
Lower: Ratio of data to SM prediction, with the filled band representing the statistical uncertainty on the data and the estimated backgrounds and the systematic uncertainty on the estimated backgrounds.}
\label{fig:resultsMT2}
\end{figure*}

The observed number of events in data in the SR are compared with the stacked SM background estimates as shown in Fig.~\ref{fig:results} (SF events), and summarised in Table~\ref{tab:results} for SF events and in Table~\ref{tab:resultseemm} for dielectron and dimuon events, separately.
The \mttwo shape of the stacked SM background estimates, the observed data and three signal scenarios are shown in Fig.~\ref{fig:resultsMT2}, for SF events, with all SR selection applied except the \mttwo requirement.
Applying the \mttwo requirement in the SR is greatly suppressing the \ttbar and Drell--Yan contributions.
\begin{table*}[ht!]
    \centering
    \topcaption{The predicted SM background contributions, their sum and the observed number of SF events in data.
        The yields expected for several signal scenarios are provided as a reference. The uncertainties associated with the background yields stem from statistical and systematic sources.
    The last bin is inclusive above 300\GeV.}
    \label{tab:results}
    \begin{tabular}{c c c c c}
        \hline
        $\ptmiss$ [{\GeVns}] & 100--150 & 150--225 & 225--300 & ${\geq}300$  \\ \hline
        FS bkg. &$96^{+13}_{-12}$ &$15.3^{+5.6}_{-4.5}$ &$4.4^{+3.6}_{-2.3}$ &$1.1^{+2.5}_{-1.0}$\\
        $\PZ\PZ$ &$13.5\pm1.5 $ &$9.78\pm1.19$&$2.84\pm0.56$&$1.86\pm0.12$\\
        $\PW\PZ$ &$6.04\pm1.19$ &$2.69\pm0.88$&$0.86\pm0.45$&$0.21\pm0.20$\\
        DY+jets&$2.01^{+0.39}_{-0.23}$& $0.00+0.28$ &$0.00+0.28$ &$0.00+0.28$\\
        Rare processes&$0.69\pm0.44$&$0.68\pm0.47$&$0.00+0.20$&$0.05\pm0.12$ \\[\cmsTabSkip]
        Total prediction &$118^{+13}_{-12}$ &$28.4^{+5.9}_{-4.8}$ &$7.9^{+3.7}_{-2.4}$ &$3.2^{+2.6}_{-1.1}$\\[\cmsTabSkip]
        Data &101 &31 &7 &7\\[\cmsTabSkip]
        $\mslep = 450\GeV$, $\mlsp = 20\GeV$&1.03$\pm$0.09& 2.67$\pm$0.15& 2.67$\pm$0.15& 8.09$\pm$0.26 \\
        $\mslep = 400\GeV$, $\mlsp = 20\GeV$&2.25$\pm$0.21& 5.05$\pm$0.31& 6.28$\pm$0.35& 13.0$\pm$0.50 \\
        $\mslep = 350\GeV$, $\mlsp = 20\GeV$&3.97$\pm$0.30& 10.9$\pm$0.49& 11.2$\pm$0.50& 15.9$\pm$0.59 \\
        \hline
    \end{tabular}
\end{table*}

\begin{table*}[ht!]
    \centering
    \topcaption{The predicted SM background contributions, their sum and the observed number of dielectron (upper) and dimuon (lower) events in data.
    The uncertainties associated with the yields stem from statistical and systematic sources. The last bin is inclusive above 300\GeV.}
    \label{tab:resultseemm}
    \bgroup
    \renewcommand*{\arraystretch}{1.2}
    \begin{tabular}{c c c c c}
        \hline
        \multicolumn{5}{c}{Dielectron events} \\
        $\ptmiss$ [{\GeVns}]& 100--150 & 150--225 & 225--300 & ${\geq}300$  \\ \hline
        FS bkg.&$36.1^{+6.6}_{-6.3}$ &$5.7^{+2.5}_{-2.1}$ &$1.6^{+1.5}_{-1.1}$ &$0.41^{+1}_{-0.5}$\\
        $\PZ\PZ$ &$5.17\pm0.68$&$3.79\pm0.58$&$1.18\pm0.31$&$0.69\pm0.07$\\
        $\PW\PZ$ &$2.65\pm0.68$&$1.16\pm0.45$&$0.39\pm0.33$&$0.21\pm0.20$\\
        DY+jets&$0.98^{+0.14}_{-0.15}$&$0.00+0.28$&$0.00+0.28$&$0.00+0.28$\\
        Rare processes&$0.02\pm0.14$&$0.26\pm0.21$&$0.00+0.11$&$0.06\pm0.04$ \\[\cmsTabSkip]
        Total prediction &$45^{+6.7}_{-6.4}$ &$11.0^{+2.6}_{-2.3}$ &$3.2^{+1.6}_{-1.2}$ &$1.4^{+1.1}_{-0.6}$\\[\cmsTabSkip]
        Data &45 &10 &2 &2\\
    \end{tabular}
    \begin{tabular}{c c c c c}
        \hline
        \multicolumn{5}{c}{Dimuon events} \\
        $\ptmiss$ [{\GeVns}]& 100--150& 150--225 & 225--300& ${\geq}300$  \\ \hline
        FS bkg.&$61.3^{+9.1}_{-8.5}$ &$9.8^{+3.9}_{-3.2}$ &$2.8^{+2.4}_{-1.7}$ &$0.70^{+1.7}_{-0.8}$\\
        $\PZ\PZ$ &$8.33\pm0.99$&$5.98\pm0.80$&$1.67\pm0.42$&$1.17\pm0.10$\\
        $\PW\PZ$ &$3.40\pm0.91$&$1.53\pm0.73$&$0.47\pm0.30$&$0.00+0.06$\\
        DY+jets&$1.03^{+0.33}_{-0.14}$ &$0.00+0.28$&$0.00+0.28$&$0.00+0.28$\\
        Rare processes&$0.66\pm0.41$&$0.42\pm0.35$&$0.00+0.16$&$0.00+0.11$ \\[\cmsTabSkip]
        Total prediction &$75^{+9.2}_{-8.7}$ &$17.7^{+4.1}_{-3.4}$ &$4.8^{+2.5}_{-1.8}$ &$1.9^{+1.7}_{-0.8}$\\[\cmsTabSkip]
        Data &56 &21 &5 &5\\
        \hline
    \end{tabular}
    \egroup
\end{table*}

At high \ptmiss values, the uncertainties in the background prediction are driven by the statistical uncertainty
in the number of events in the DF sample used to derive the FS background.
There is agreement between observation and SM expectation given the systematic and statistical uncertainties.

\section{Interpretation}
\label{sec:interpretation}

The results are interpreted in terms of the simplified model described in Section~\ref{sec:introduction}.
Upper limits on the cross section, assuming branching ratios of 100\%, have been calculated at 95\% confidence level (\CL) using the \CLs criterion
and an asymptotic formulation~\cite{Junk:1999kv,0954-3899-28-10-313,HiggsTool1,Cowan:2010js},
taking into account the statistical and systematic uncertainties in the signal yields and the background predictions.
Systematic uncertainties are modelled using a log-normal distribution in the fit, except for the FS uncertainty where the distribution of the nuisance parameter is modelled using the gamma function.
The uncertainties are correlated among the SRs and are considered as such in a combined fit to all SRs simultaneously.
\subsection{Systematic uncertainty in the signal yield}
The systematic uncertainties associated with the signal are shown in Table~\ref{tab:systs}, and described below.
The uncertainty in the measurement of the integrated luminosity is 2.5\%~\cite{CMS-PAS-LUM-17-001}.
A flat uncertainty of 5\% is associated to the lepton identification and isolation efficiency in the signal acceptance~\cite{Sirunyan:2018fpa,Khachatryan:2015hwa}, an uncertainty of up to 2.5\,(3)\% in the electron (muon) efficiency of the signal using fast simulation, and dedicated corrections for fast simulation to match the data are applied.
The uncertainty on the trigger efficiency is measured to be 3\%, as described in Section~\ref{sub:fsbkg}.
The uncertainty in the jet energy scale is assessed by shifting the jet energy correction factors for each jet by one standard deviation up and down and recalculating
the kinematic quantities. The result varies between 1 and 16\% depending on the signal kinematics.
The uncertainty due to the simulation of pileup for simulated background processes is taken into account by varying the expected cross section of inelastic collisions by 5\%~\cite{Sirunyan:2018nqx}, and amounts to 0.5--7\% depending on the signal scenario.
Varying the unclustered energy, and the electron and muon scales, up and down by their $1\sigma$ variations and evaluating the effect on the \ptmiss, results in systematic
uncertainties of 0.5--8\%, 0.5--4\%, and 0.5--20\% respectively.
The large variation associated to the muon energy scale is due to the poor muon momentum resolution at high \pt, and the quoted value is driven by one signal scenario containing such high \pt muon events.
The theoretical uncertainties are those related to the uncertainty on the QCD renormalisation ($\mu_{\text{R}}$) and factorisation ($\mu_{\text{F}}$) scales, and of the PDF.
The systematic uncertainties associated with the $\mu_{\text{R}}$ and $\mu_{\text{F}}$ scales are evaluated using weights derived from the SysCalc code applied to simulated signal
events~\cite{Kalogeropoulos:2018cke}.
For renormalisation and factorisation scales the Stewart--Tackmann prescription~\cite{PhysRevD.85.034011} is followed, that treats the theory uncertainties in analyses with a jet selection.
This procedure results in an uncertainty of 1--11\%.
Finally the statistical uncertainty in the number of simulated events is also considered and found
to be in the range 0.5--20\%, depending on the signal scenario.
\begin{table}[htb]
\centering
\topcaption{\label{tab:systs} List of systematic uncertainties taken into account for the signal yields.}
\begin{tabular}{l c c}
\hline
Source of uncertainty                   & Uncertainty (\%)\\
\hline
Integrated luminosity                   &      2.5      \\
Lepton reconstruction/isolation eff.    &      5        \\
Trigger modelling                        &      3        \\
Fast simulation electron efficiency     &      1--2.5   \\
Fast simulation muon efficiency         &      1--3     \\
Jet energy scale                        &      1--15    \\
Pileup                                  &      0.5--7   \\
Fast simulation \ptmiss modelling        &      0.5--20  \\
Unclustered energy shifted \ptmiss      &      0.5--8   \\
Muon energy scale shifted \ptmiss       &      0.5--20  \\
Electron energy scale shifted \ptmiss   &      0.5--4   \\
Renormalisation/factorisation scales  	&      1--11    \\
PDF                                  	&      3        \\
MC statistical uncertainty              &      0.5--20  \\ \hline
\end{tabular}
\end{table}

\subsection{Interpretations using simplified models}
\begin{figure*}[htbp]
\centering
\includegraphics[width=0.48\textwidth]{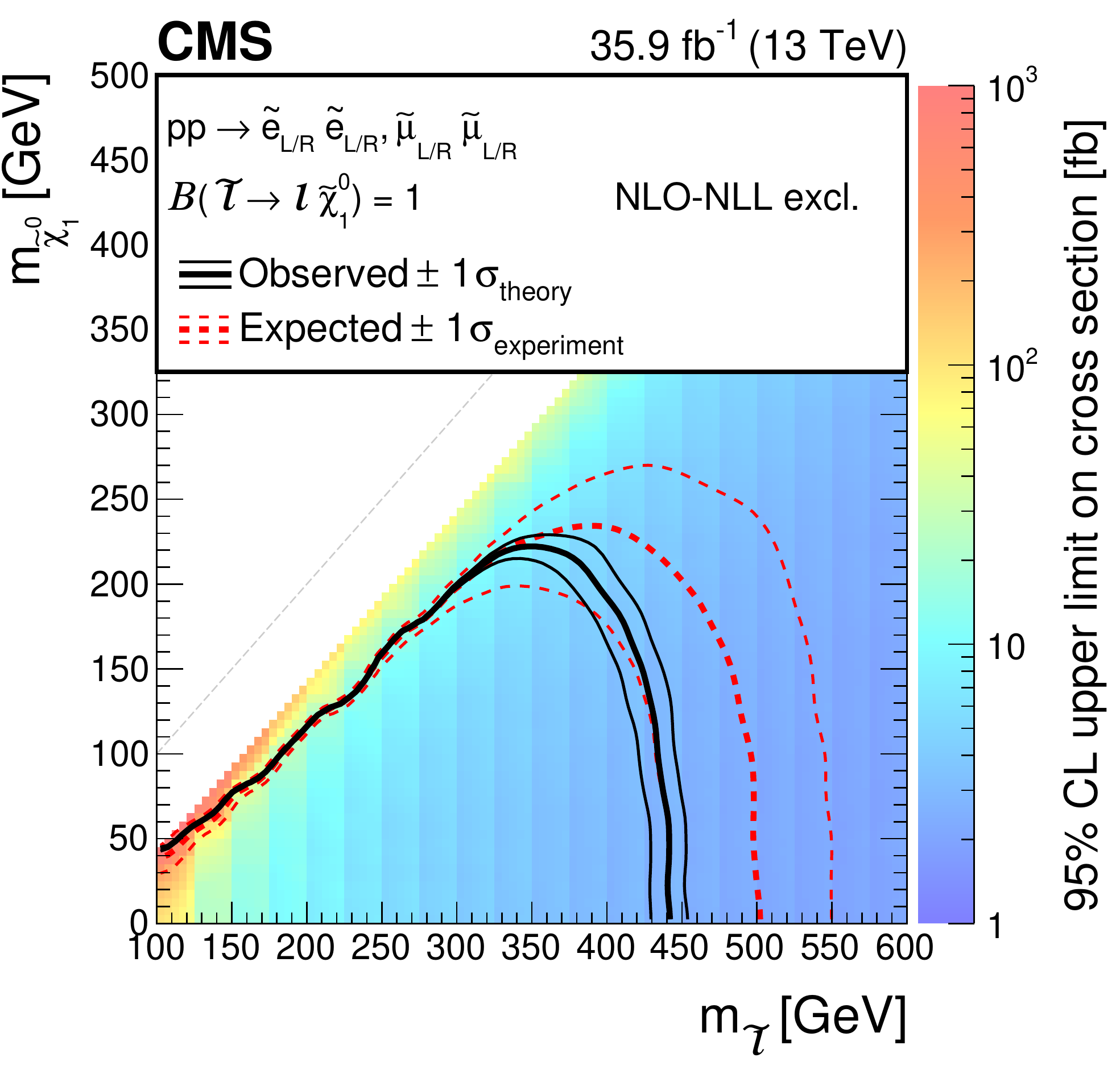} \\
\includegraphics[width=0.48\textwidth]{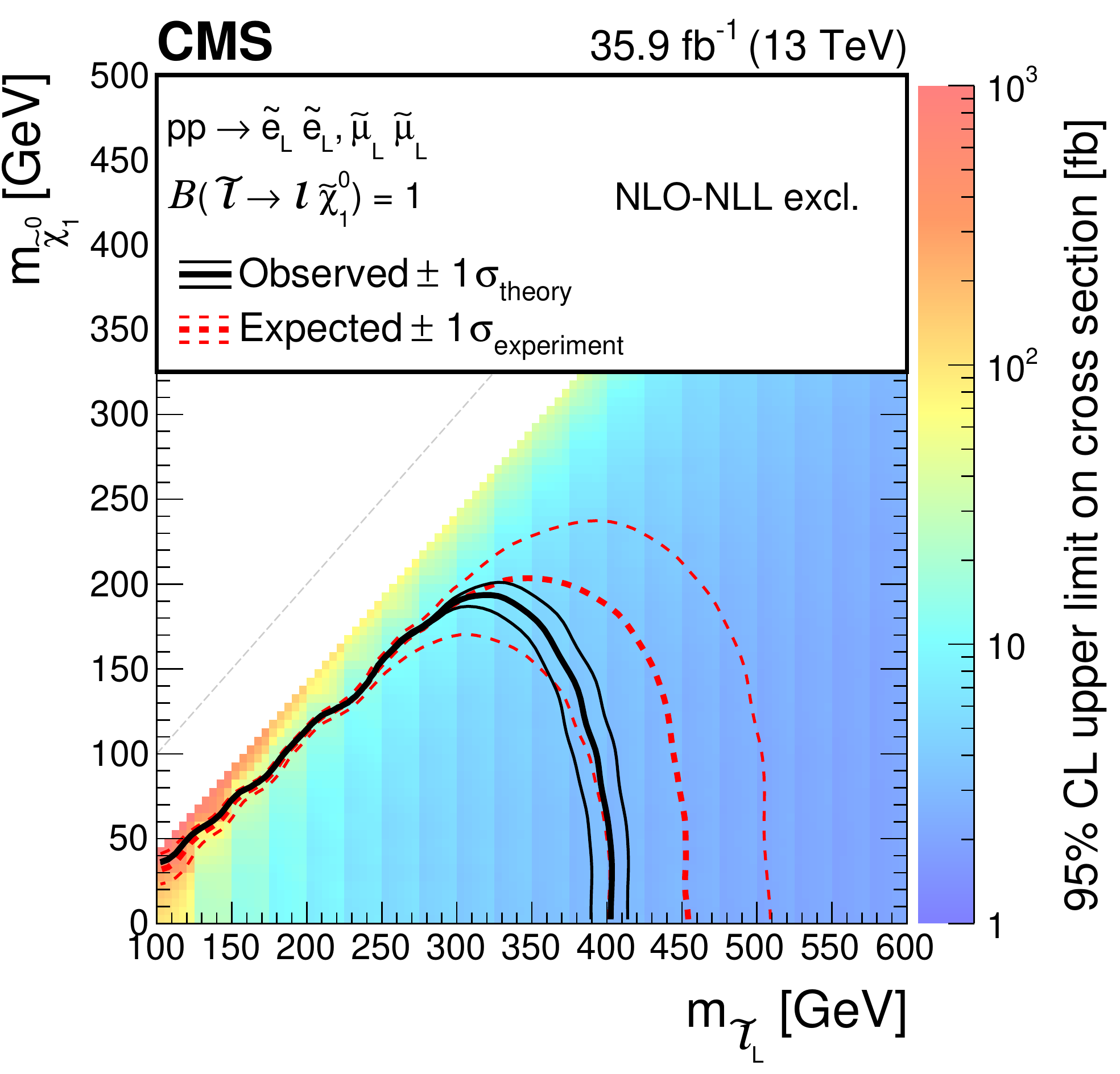}
\includegraphics[width=0.48\textwidth]{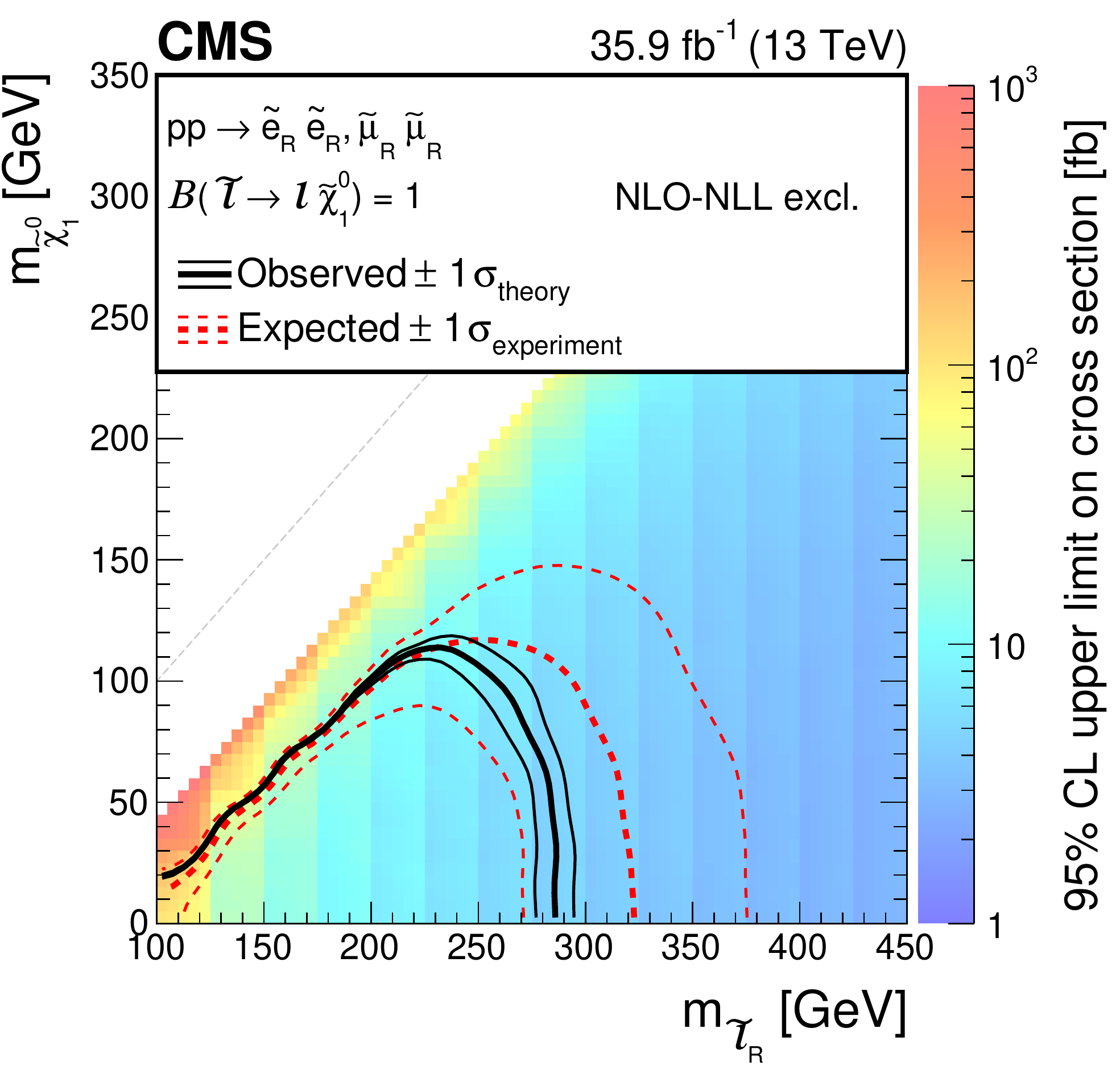}
\caption{\label{fig:TOT}
Cross section upper limit and exclusion contours at 95\% \CL for direct slepton production of two flavours, selectrons and smuons,
as a function of the \lsp and \slep masses, assuming the production of both left- and right-handed sleptons
(upper) or production of only left- (lower left) or right-handed (lower right).
The region under the thick red dotted (black solid) line is excluded by the expected (observed) limit.
The thin red dotted curves indicate the regions containing 95\% of the distribution of limits
expected under the background-only hypothesis.
The thin solid black curves show the change in the observed limit due to
variation of the signal cross sections within their theoretical uncertainties.
}
\end{figure*}
\begin{figure*}[htbp]
\centering
\includegraphics[width=0.48\textwidth]{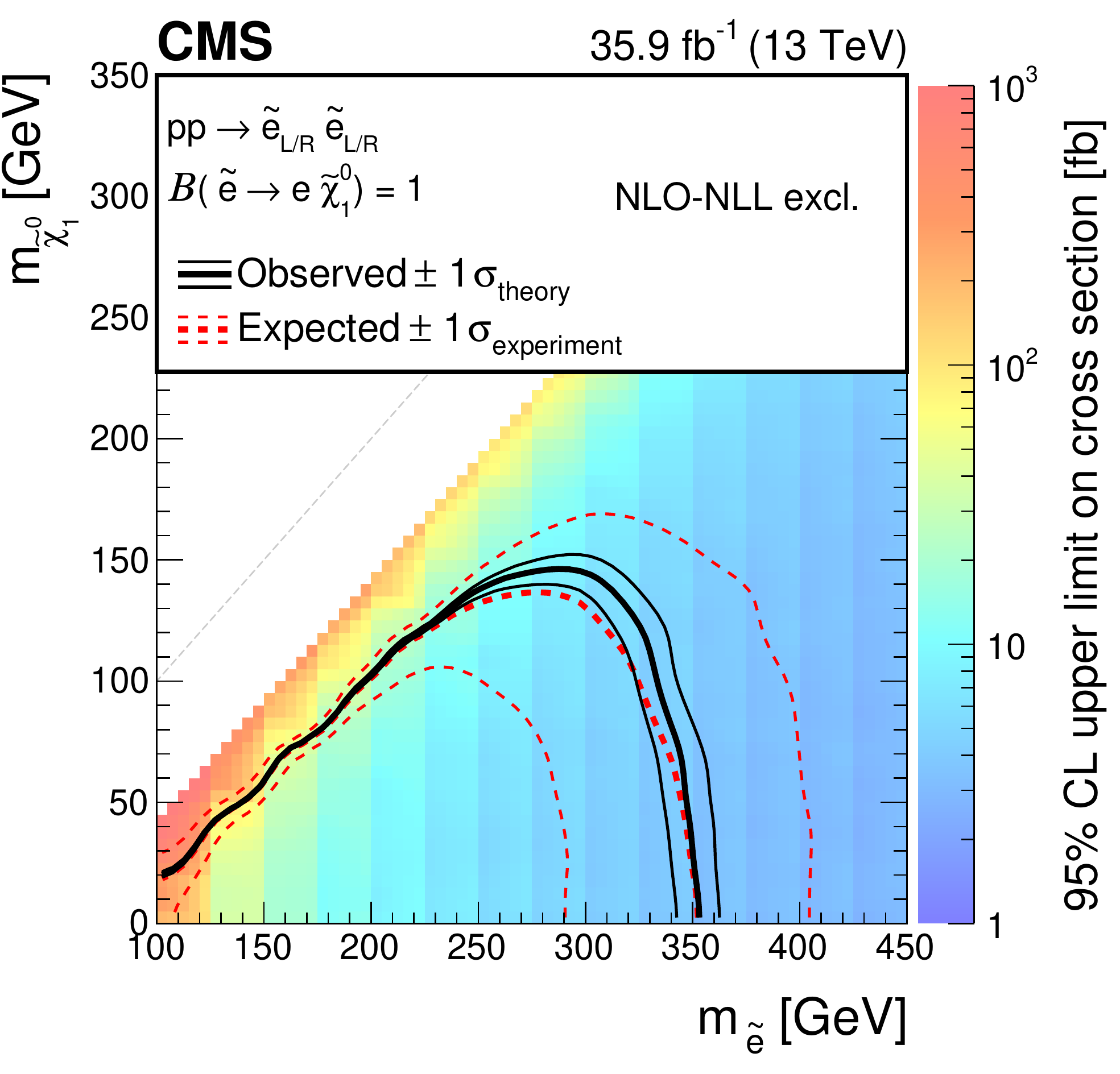} \\
\includegraphics[width=0.48\textwidth]{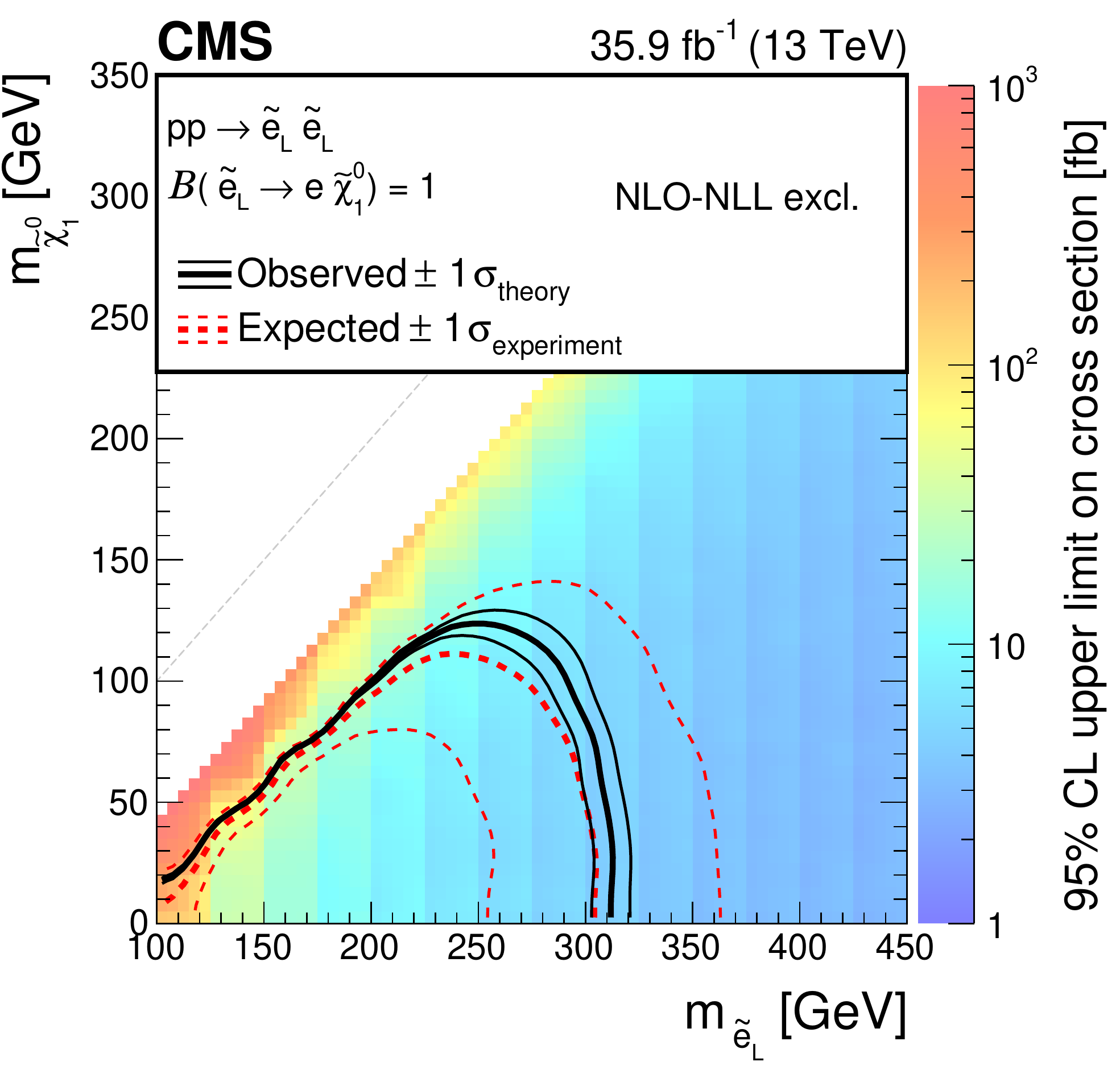}
\includegraphics[width=0.48\textwidth]{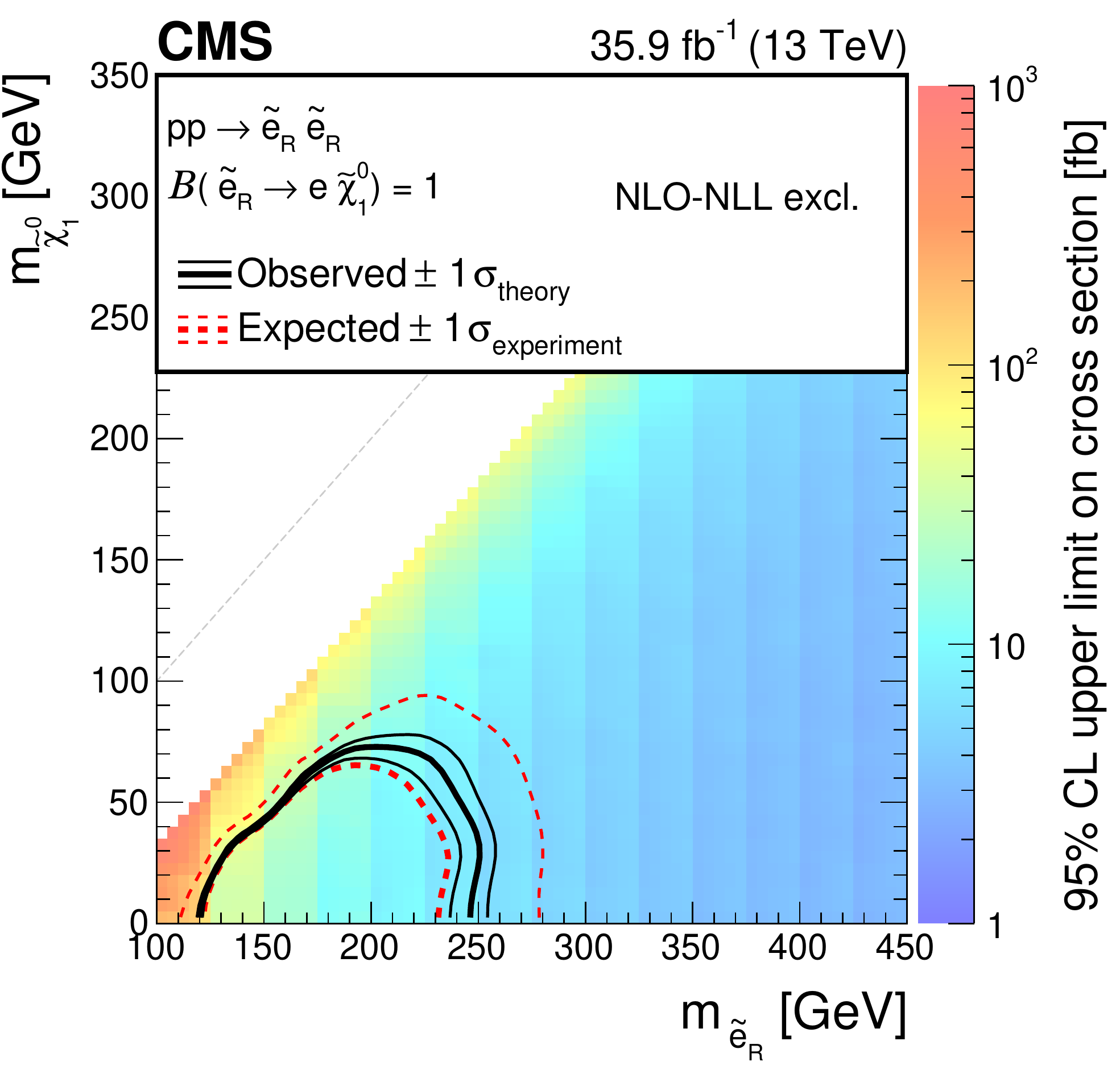}
\caption{\label{fig:TOTee}
Cross section upper limit and exclusion contours at 95\% \CL for direct selectron production
as a function of the \lsp and \slep masses, assuming the production of both left- and right-handed selectrons
(upper), or production of only left- (lower left) or right-handed (lower right) selectrons.
The region under the thick red dotted (black solid) line is excluded by the expected (observed) limit.
The thin red dotted curves indicate the regions containing 95\% of the distribution of limits
expected under the background-only hypothesis. For the right-handed selectrons, only the $+1\sigma$ expected line (thin red dotted curve) is shown as no exclusion can be made at $-1\sigma$.
The thin solid black curves show the change in the observed limit due to
variation of the signal cross sections within their theoretical uncertainties.
}
\end{figure*}
 \begin{figure*}[htbp]
\centering
\includegraphics[width=0.48\textwidth]{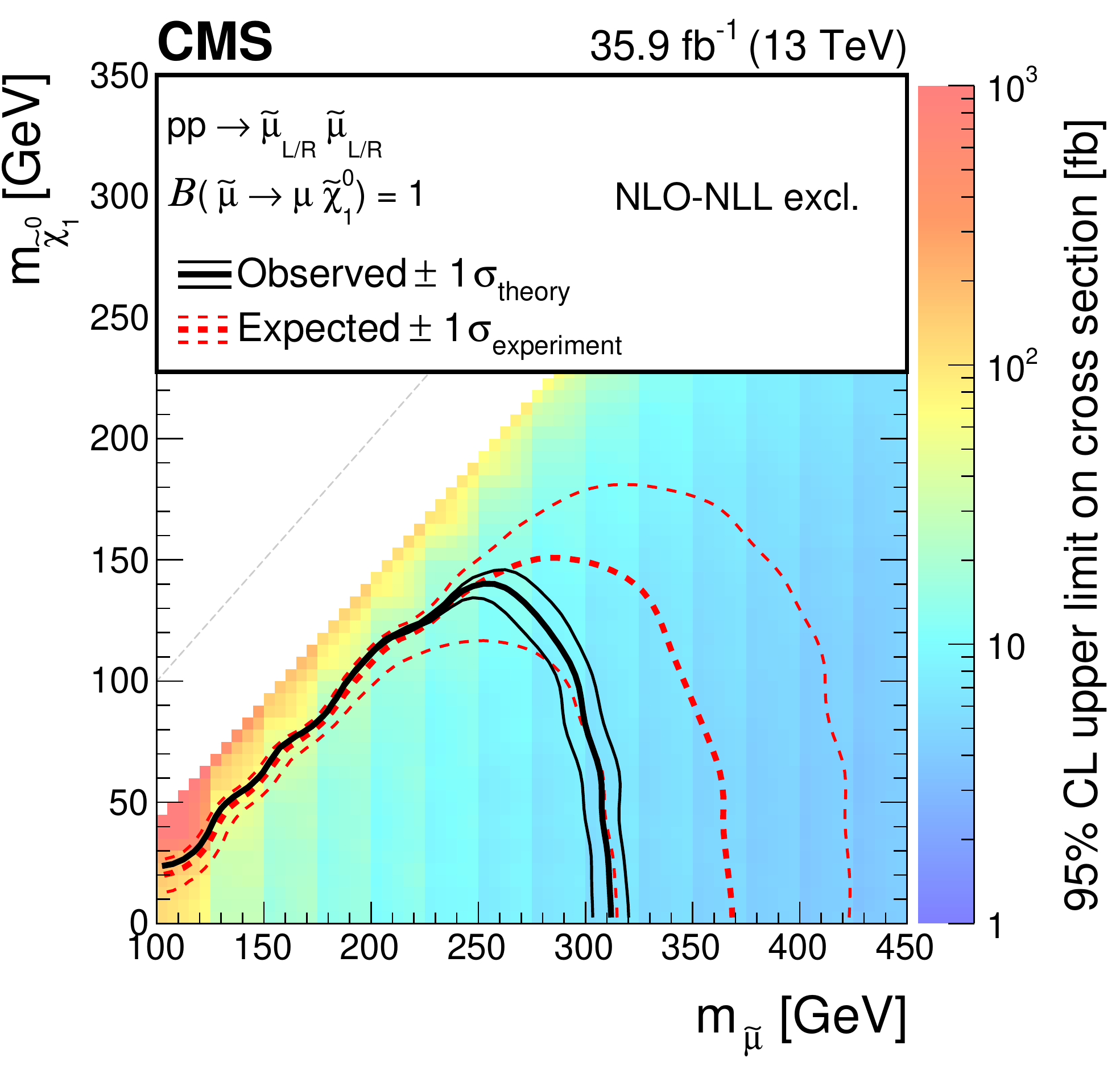} \\
\includegraphics[width=0.48\textwidth]{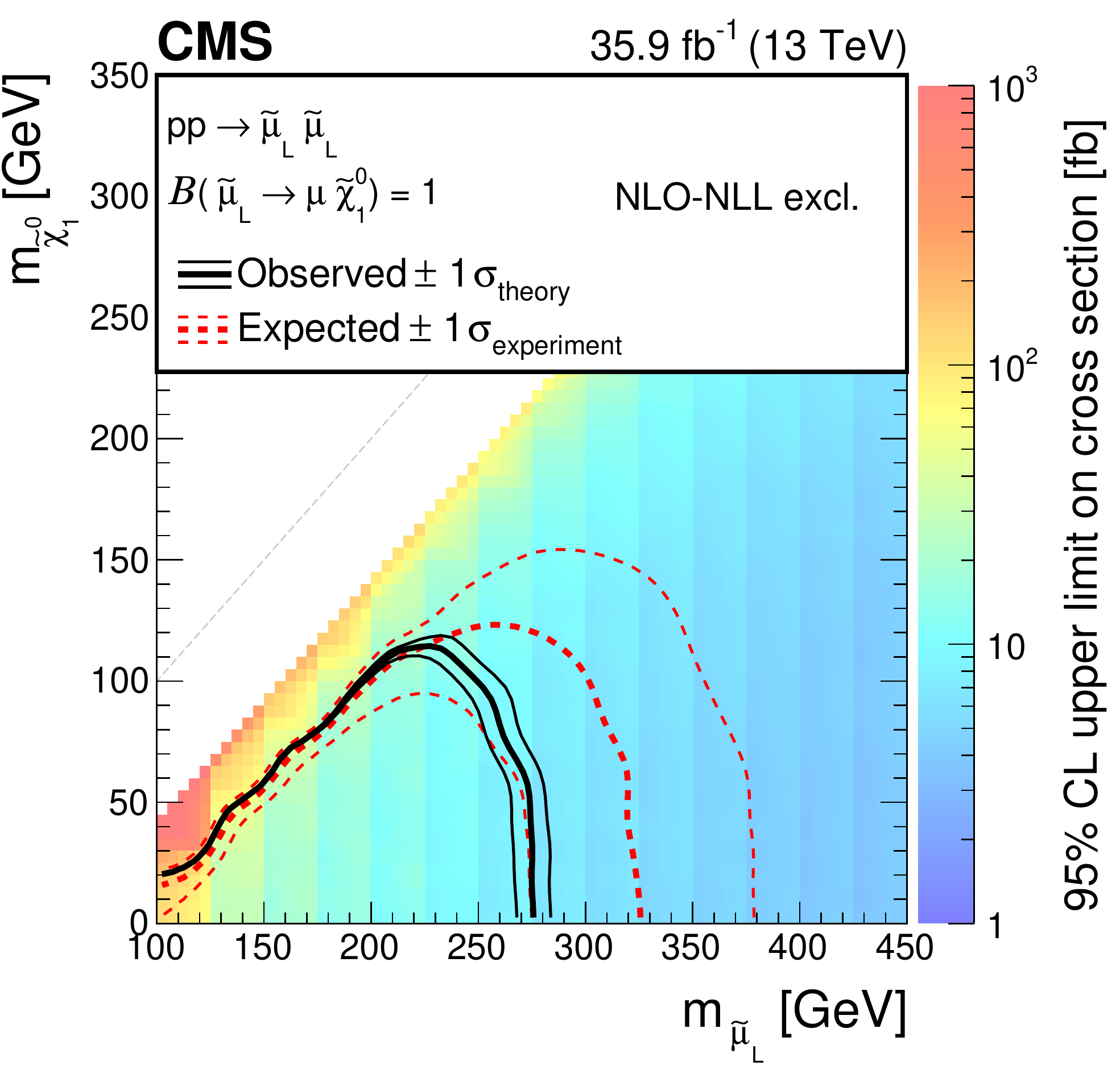}
\includegraphics[width=0.48\textwidth]{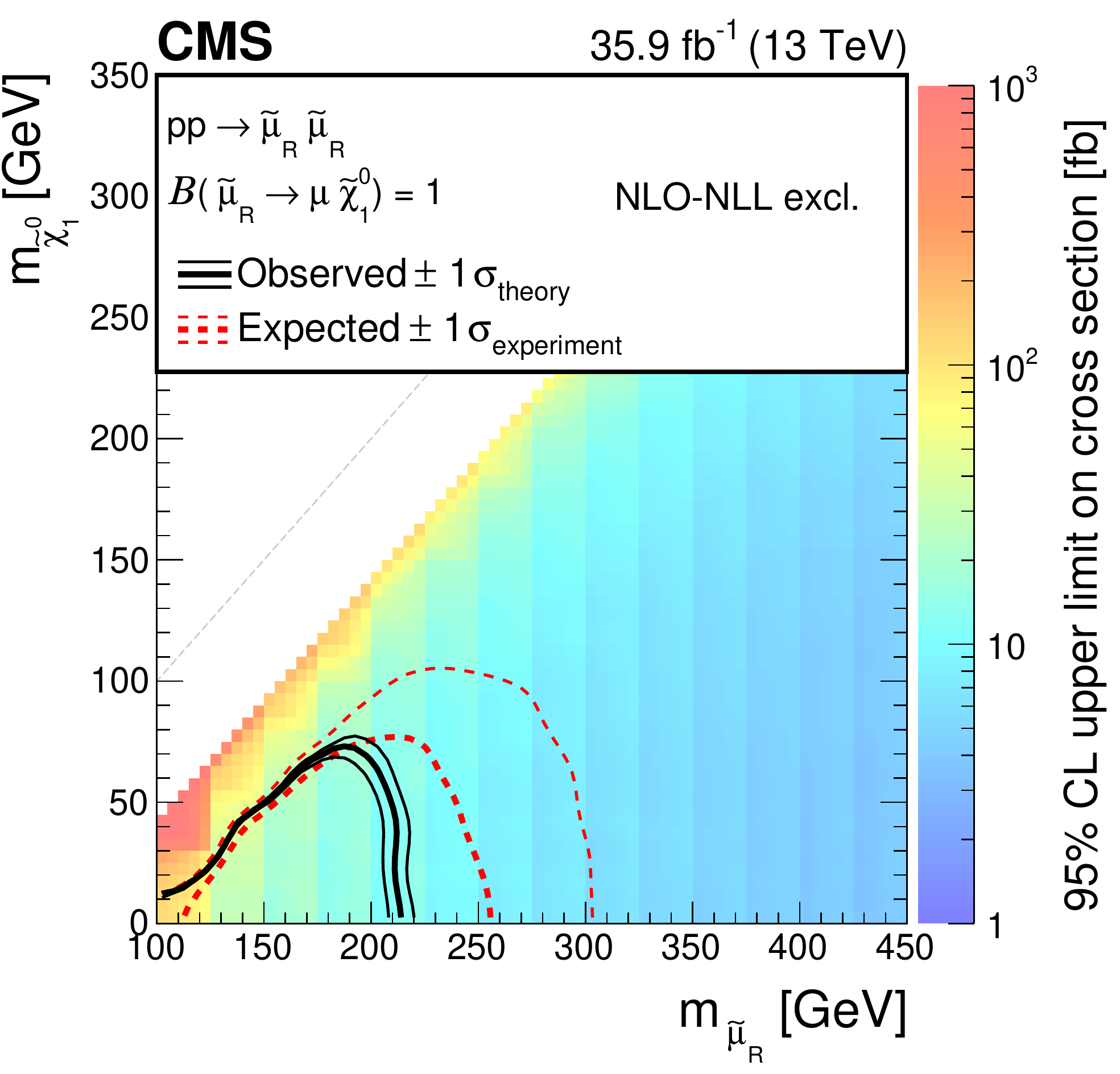}
\caption{\label{fig:TOTmm}
Cross section upper limit and exclusion contours at 95\% \CL for direct smuon production
as a function of the \lsp and \slep masses, assuming the production of both left- and right-handed smuons
(upper), or production of only left- (lower left) or right-handed (lower right) smuons.
The region under the thick red dotted (black solid) line is excluded by the expected (observed) limit.
The thin red dotted curves indicate the regions containing 95\% of the distribution of limits
expected under the background-only hypothesis. For the right-handed smuons, only the $+1\sigma$ expected line (thin red dotted curve) is shown as no exclusion can be made at $-1\sigma$.
The thin solid black curves show the change in the observed limit due to
variation of the signal cross sections within their theoretical uncertainties.
}
\end{figure*}

Upper limits on the direct slepton pair production cross section are displayed in Fig.~\ref{fig:TOT} for three scenarios: assuming the existence of both flavour mass degenerate left- and
right-handed sleptons, for only left-handed sleptons, and for only right-handed sleptons. Similarly, the limits on direct selectron and smuon production are displayed in Figs.~\ref{fig:TOTee}
and~\ref{fig:TOTmm}, respectively.
The Figs.~\ref{fig:TOT}-\ref{fig:TOTmm} also show the $95\%$ \CL exclusion contours,
as a function of the \slep and \lsp masses.
Note that the cross section at a given mass for right-handed sleptons is expected to be about one third of that for left-handed sleptons.
The analysis probes slepton masses up to approximately 450, 400, or 290\GeV, assuming both left- and right-handed, left-handed only, or right-handed sleptons, and a massless LSP.
For models with high slepton masses and light LSPs the sensitivity is driven by the highest \ptmiss bin. The sensitivity is reduced at higher LSP masses due to the effect of the lepton acceptance.
In the case of selectrons (smuons), the limits corresponding to these 3 scenarios are 350, 310 and 250\GeV (310, 280, and 210\GeV).
Since the dimuon data yield in the highest \ptmiss\ bin is somewhat higher than predicted, the observed limits in this channel are weaker than expected in the absence of signal.
These results improve the previous 8\TeV exclusion limits by 100--150\GeV in the slepton mass~\cite{2012ewk}.
\section{Summary}
\label{sec:summary}
A search for direct slepton (selectron or smuon) production, in events with opposite-charge, same-flavour leptons,
no jets, and missing transverse momentum has been presented.
The data comprise a sample of proton-proton collisions collected with the CMS detector in
2016 at a centre-of-mass energy of 13\TeV, corresponding to an integrated luminosity of \lint.
Observations are in agreement with Standard Model expectations within the
statistical and systematic uncertainties.
Exclusion limits are provided assuming right-handed only, left-handed only and right-and left-handed two flavour slepton production scenarios (mass degenerate selectrons and smuons).
Slepton masses up to 290, 400 and 450\GeV respectively are excluded at 95\% confidence level, assuming a massless LSP. Exclusion limits are also provided assuming a massless LSP and right-handed
only, left-handed only and right-and left-handed single flavour production scenarios, excluding selectron (smuon) masses up to 250, 310 and 350\GeV (210, 280 and 310\GeV), respectively.
These results improve the previous exclusion limits measured by the CMS experiment at a centre-of-mass energy of 8\TeV by 100–-150\GeV in slepton masses.

\begin{acknowledgments}
We congratulate our colleagues in the CERN accelerator departments for the excellent performance of the LHC and thank the technical and administrative staffs at CERN and at other CMS institutes for their contributions to the success of the CMS effort. In addition, we gratefully acknowledge the computing centres and personnel of the Worldwide LHC Computing Grid for delivering so effectively the computing infrastructure essential to our analyses. Finally, we acknowledge the enduring support for the construction and operation of the LHC and the CMS detector provided by the following funding agencies: BMWFW and FWF (Austria); FNRS and FWO (Belgium); CNPq, CAPES, FAPERJ, FAPERGS, and FAPESP (Brazil); MES (Bulgaria); CERN; CAS, MOST, and NSFC (China); COLCIENCIAS (Colombia); MSES and CSF (Croatia); RPF (Cyprus); SENESCYT (Ecuador); MoER, ERC IUT, and ERDF (Estonia); Academy of Finland, MEC, and HIP (Finland); CEA and CNRS/IN2P3 (France); BMBF, DFG, and HGF (Germany); GSRT (Greece); NKFIA (Hungary); DAE and DST (India); IPM (Iran); SFI (Ireland); INFN (Italy); MSIP and NRF (Republic of Korea); LAS (Lithuania); MOE and UM (Malaysia); BUAP, CINVESTAV, CONACYT, LNS, SEP, and UASLP-FAI (Mexico); MBIE (New Zealand); PAEC (Pakistan); MSHE and NSC (Poland); FCT (Portugal); JINR (Dubna); MON, ROSATOM, RAS, RFBR, and NRC KI (Russia); MESTD (Serbia); SEIDI, CPAN, PCTI, and FEDER (Spain); Swiss Funding Agencies (Switzerland); MST (Taipei); ThEPCenter, IPST, STAR, and NSTDA (Thailand); TUBITAK and TAEK (Turkey); NASU and SFFR (Ukraine); STFC (United Kingdom); DOE and NSF (USA).
\hyphenation{Rachada-pisek} Individuals have received support from the Marie-Curie program and the European Research Council and Horizon 2020 Grant, contract No. 675440 (European Union); the Leventis Foundation; the Alfred P. Sloan Foundation; the Alexander von Humboldt Foundation; the Belgian Federal Science Policy Office; the Fonds pour la Formation \`a la Recherche dans l'Industrie et dans l'Agriculture (FRIA-Belgium); the Agentschap voor Innovatie door Wetenschap en Technologie (IWT-Belgium); the F.R.S.-FNRS and FWO (Belgium) under the ``Excellence of Science - EOS" - be.h project n. 30820817; the Ministry of Education, Youth and Sports (MEYS) of the Czech Republic; the Lend\"ulet (``Momentum") Program and the J\'anos Bolyai Research Scholarship of the Hungarian Academy of Sciences, the New National Excellence Program \'UNKP, the NKFIA research grants 123842, 123959, 124845, 124850 and 125105 (Hungary); the Council of Science and Industrial Research, India; the HOMING PLUS program of the Foundation for Polish Science, cofinanced from European Union, Regional Development Fund, the Mobility Plus program of the Ministry of Science and Higher Education, the National Science Centre (Poland), contracts Harmonia 2014/14/M/ST2/00428, Opus 2014/13/B/ST2/02543, 2014/15/B/ST2/03998, and 2015/19/B/ST2/02861, Sonata-bis 2012/07/E/ST2/01406; the National Priorities Research Program by Qatar National Research Fund; the Programa Estatal de Fomento de la Investigaci{\'o}n Cient{\'i}fica y T{\'e}cnica de Excelencia Mar\'{\i}a de Maeztu, grant MDM-2015-0509 and the Programa Severo Ochoa del Principado de Asturias; the Thalis and Aristeia programs cofinanced by EU-ESF and the Greek NSRF; the Rachadapisek Sompot Fund for Postdoctoral Fellowship, Chulalongkorn University and the Chulalongkorn Academic into Its 2nd Century Project Advancement Project (Thailand); the Welch Foundation, contract C-1845; and the Weston Havens Foundation (USA).

\end{acknowledgments}

\bibliography{auto_generated}

\cleardoublepage \appendix\section{The CMS Collaboration \label{app:collab}}\begin{sloppypar}\hyphenpenalty=5000\widowpenalty=500\clubpenalty=5000\vskip\cmsinstskip
\textbf{Yerevan Physics Institute, Yerevan, Armenia}\\*[0pt]
A.M.~Sirunyan, A.~Tumasyan
\vskip\cmsinstskip
\textbf{Institut f\"{u}r Hochenergiephysik, Wien, Austria}\\*[0pt]
W.~Adam, F.~Ambrogi, E.~Asilar, T.~Bergauer, J.~Brandstetter, E.~Brondolin, M.~Dragicevic, J.~Er\"{o}, A.~Escalante~Del~Valle, M.~Flechl, R.~Fr\"{u}hwirth\cmsAuthorMark{1}, V.M.~Ghete, J.~Hrubec, M.~Jeitler\cmsAuthorMark{1}, N.~Krammer, I.~Kr\"{a}tschmer, D.~Liko, T.~Madlener, I.~Mikulec, N.~Rad, H.~Rohringer, J.~Schieck\cmsAuthorMark{1}, R.~Sch\"{o}fbeck, M.~Spanring, D.~Spitzbart, A.~Taurok, W.~Waltenberger, J.~Wittmann, C.-E.~Wulz\cmsAuthorMark{1}, M.~Zarucki
% ...
\vskip\cmsinstskip
% ...
1:  Also at Vienna University of Technology, Vienna, Austria\\
% ...
\end{sloppypar}
\end{document}